\documentclass[%
 reprint,
 amsmath,amssymb,
 aps,
,onecolumn, notitlepage]{revtex4-1}

\pdfoutput=1
\usepackage[table]{xcolor}
\usepackage{multirow}
\usepackage[percent]{overpic}
\usepackage[utf8]{inputenc} 
\usepackage[T1]{fontenc}    
\usepackage{hyperref}       
\usepackage{url}            
\usepackage{booktabs}       
\usepackage{amsfonts}       
\usepackage{nicefrac}       
\usepackage{microtype}      
\usepackage[section]{placeins}
\usepackage{changepage}
\usepackage{amsmath,amsfonts,amsthm,bm}
\usepackage{pgfplots}
\usepackage{float}
\usepackage{tikz}
\usepackage{setspace}
\usepackage{amssymb}
\usepackage{dcolumn}
\usepackage{siunitx}
\usepgfplotslibrary{groupplots}
\usetikzlibrary{calc,decorations.markings}
\PassOptionsToPackage{dvipsnames,svgnames}{xcolor}   
\usetikzlibrary{arrows,calc}
 \makeatletter%
\usepackage[toc,page]{appendix}
\usetikzlibrary{3d,calc}
\makeatletter
\newcommand{\vast}{\bBigg@{4}}
\newcommand{\Vast}{\bBigg@{5}}
\makeatother
\pgfplotsset{compat=1.14} 

\usetikzlibrary{shapes.misc}
\tikzset{cross/.style={cross out, draw=black, minimum size=2*(#1-\pgflinewidth), inner sep=0pt, outer sep=0pt},
cross/.default={3pt}}

\begin{document}

\preprint{APS Preprint}

\title{Optimal Inverse Design of Magnetic Field Profiles in a Magnetically Shielded Cylinder}

\author{M.~Packer\textsuperscript{1}}
\author{P.~J.~Hobson\textsuperscript{1}}%
\author{N.~Holmes\textsuperscript{2}}%
\author{J.~Leggett\textsuperscript{2}}%
\author{P.~Glover\textsuperscript{2}}%
\author{M.~J.~Brookes\textsuperscript{2}}%
\author{R.~Bowtell\textsuperscript{2}}%
\author{T.~M.~Fromhold\textsuperscript{1}}%
\affiliation{%
 \textsuperscript{1}School of Physics and Astronomy, University of Nottingham, Nottingham, NG7 2RD, UK. \\
 \textsuperscript{2}Sir Peter Mansfield Imaging Centre, School of Physics and Astronomy, University of Nottingham, Nottingham, NG7 2RD, UK.
}
\date{\today}
\begin{abstract}
Magnetic shields that use both active and passive components to enable the generation of a tailored low-field environment are required for many applications in science, engineering, and medical imaging. Until now, accurate field nulling, or field generation, has only been possible over a small fraction of the overall volume of the shield. This is due to the interaction between the active field-generating components and the surrounding high-permeability passive shielding material. In this paper, we formulate the interaction between an arbitrary static current flow on a cylinder and an exterior closed high-permeability cylinder. We modify the Green's function for the magnetic vector potential and match boundary conditions on the shield's interior surface to calculate the total magnetic field generated by the system. We cast this formulation into an inverse optimization problem to design active--passive magnetic field shaping systems that accurately generate any physical static magnetic field in the interior of a closed cylindrical passive shield. We illustrate this method by designing hybrid systems that generate a range of magnetic field profiles to high accuracy over large interior volumes, and simulate them in real-world shields whose passive components have finite permeability, thickness, and axial entry holes. Our optimization procedure can be adapted to design active--passive magnetic field shaping systems that accurately generate any physical user-specified static magnetic field in the interior of a closed cylindrical shield of any length, enabling the development and miniaturization of systems that require accurate magnetic shielding and control.
\end{abstract}

\maketitle

\section{Introduction}
Regions of space that require precisely-controlled magnetic field environments are essential for a wide range of experiments, devices, and applications. These include quantum sensing of gravity and gravity gradients for geophysical survey and mapping~\cite{Wueaax0800, doi:10.1038/s41598-018-30608-1, SnaddenGradiometer, AIQuantumSensors}; atomic magnetometry~\cite{PhysRevLett.113.013001,quspin,10.1155/2015/491746} for applications including material characterization~\cite{ROMALIS2011258} and magnetoencephalography~\cite{10.1364/BOE.3.000981, nature}; noise reduction in fundamental physics experiments~\cite{doi:10.1063/1.4919366, 10.1007/s10751-014-1109-5} such as timing using high-precision atomic clocks~\cite{MORIC2014287, LiangClock}, measuring the electric dipole moment of fundamental systems~\cite{RevModPhys.62.541, HindsEDM, ACMEEDM, doi:10.1063/1.4922671}, and testing Lorentz-invariance by observing the limits of spin precession~\cite{PhysRevLett.105.151604, PhysRevA.100.010501, PhysRevLett.112.110801}. Typically, these systems are enclosed within high-permeability materials, such as mumetal, to shield magnetically sensitive components from undesired magnetic fields. Cylindrical geometries, in particular, are widely used due to their comparatively high shielding factor relative to their low manufacturing cost~\cite{10.1063/1.1656455, GRABCHIKOV201649, 6217348}.\\

Romeo and Hoult first laid the mathematical framework for the design of magnetic fields in free space through the use of current loops and arcs~\cite{RandH}. These simple elements were expanded in a spherical harmonic basis and unwanted field profiles removed by adjusting the relative separations of the discrete coils. Inverse methods based on formulating a current density in terms of simple triangular boundary elements~\cite{Pissanetzky_1992,doi:10.1002/cmr.b.20091,doi:10.1002/cmr.b.20040} enabled the design of coils of arbitrary complexity, giving greater flexibility in the design of systems at the cost of computational power and numerical instability. Pseudo-analytical formulations using a Green’s function expansion and harmonic minimization methods have been developed to generate accurate user-specified magnetic fields in free space~\cite{Forbes_2002,Forbes_2001,niall0} to facilitate the need for fast and accurate design procedures. However, these methods did not incorporate the interaction of high-permeability passive shielding material, hindering the accurate generation of specified target magnetic field profiles~\cite{doi:10.1002/9780470268483.app2} in shielded environments. Consequently, optimization of magnetic field cancellation, or other field shaping, systems in the presence of a material with high magnetic permeability is a long-standing challenge in electromagnetism.\\

Finite element methods (FEMs) can be used to develop models of hybridized active and passive shielding systems. One method of optimizing active--passive systems using FEMs would be to use genetic algorithms~\cite{genetic, genetic1} to evolve the coil parameters iteratively, including the effect of the passive material on the magnetic field generated actively at each iteration. Due to their computational complexity, however, FEMs can be slow, and, if coupled with forward optimization procedures provide only locally optimal designs since desired magnetic field profiles depend highly non-linearly on the system parameters. Analytical formulations of the magnetic field generated by hybrid active--passive systems have the distinct advantage that optimal solutions can be calculated at a range of target points with minimal computation~\cite{CACIAGLI2018423}. Currently, analytical solutions for coils in high-permeability cylindrical magnetic shields have only been formulated in the specific cases of magnetically-shielded solenoids and discrete current loops~\cite{Solenoid1, solenoid2, doi:10.1063/1.1719514}. The geometric parameters of the active structure in these systems, such as the separation distances of wire loops, are adapted to account for the interaction between the active and passive systems. Previously, the method of mirror images~\cite{jackson} has been used to formulate the response of a planar material with high magnetic permeability to a magnetic field generated by a current source~\cite{cubefem, niall, 8500866, LIU2020166846}. In these formulations, Maxwell's equations are solved explicitly by including the reflections of the current source generated by the high-permeability material in order to match the required boundary conditions. More generally, Green's function solutions to boundary value problems have been calculated for an extensive range of electromagnetic systems~\cite{greensey}, but have not previously been found generally for the case of a finite closed cylindrical high-permeability shield in the presence of an arbitrary cylindrical coaxial static current source.\\

To address this, here, we incorporate \emph{ab initio} the effect of a finite length high-permeability cylinder on the magnetic field generated by an arbitrary static current flow on a conducting cylinder, and use our results to determine the flow required to generate specified static target fields optimally. This enables the geometry of the active components on the surface of a cylinder to be determined entirely from a desired magnetic field profile. Guided by~\cite{turner}, we first derive a Green's function for a hybrid active--passive field-generating system described in cylindrical coordinates. We then utilize a modified Fourier basis to define an adjusted current density distribution which accounts for the effect of the high-permeability material. From this, we determine globally-optimal current density maps using a quadratic optimization procedure, akin to magnetic field design methodologies used previously in Magnetic Resonance Imaging (MRI)~\cite{doi:10.1002/mrm.1910260202, hoult}. To construct practical current sources from these current density maps, we define the streamfunction of the current continuum and discretize it into a set of closed-loop current-carrying wire geometries. Using this formulation, we present two example coil designs optimized in the interior of a closed cylindrical magnetic shield of finite length and high permeability, $\mu_r\gg 1$, and show that our analytical model agrees well with FEM simulations performed for the optimized current flow patterns. We then show that our design methodology can be used in a real-world situation where the cylindrical magnetic shield has finite thickness and permeability as well as axial entry holes in the end caps to allow experimental access. Using this formulation, we transform the performance of systems designed to generate user-specified magnetic field profiles in the magnetostatic regime, reducing the amount of high-permeability material required, and find globally-optimal solutions required for practical, low Size, Weight, Power, and Cost (SWaP-C) technologies for the applications listed above.

\section{Theory}\label{sec.theory}
 The interface conditions for a magnetic field along a boundary, $S$, between two materials are that the normal component of the magnetic field, $\mathbf{B}$, and tangential component of the magnetic field strength, $\textbf{H}$, are continuous. Considering the boundary between air and a material, working in the magnetostatic regime, where no eddy currents are induced, and in the case where no surface currents are present, these interface conditions are
\begin{equation}\label{eq.bperp}
    (\mathbf{B_{\mathrm{mat.}}}-\mathbf{B_{\mathrm{air}}})\cdot \mathbf{\hat{n}}=0 \qquad \textnormal{on $S$},
\end{equation}
and
\begin{equation}\label{eq.btan}
    \left(\frac{1}{\mu_r}\mathbf{B_{\mathrm{mat.}}}-\mathbf{B_{\mathrm{air}}}\right)\wedge \mathbf{\hat{n}}=0 \qquad \textnormal{on $S$},
\end{equation}
where $\hat{\mathbf{n}}$ is the unit vector normal to the boundary, and $\mu_r$ is the relative permeability of the material. In free space, the magnetic field is related to the magnetic field strength and the magnetization, $\textbf{M}$, by
\begin{equation}\label{eq.mag}
    \mathbf{B}=\mu_0 (\mathbf{H}+\textbf{M}).
\end{equation}
Physically, the magnetization of the sub-domains of a material change in response to an applied magnetic field to satisfy the boundary condition, \eqref{eq.btan}, at the material's surface. Here, we choose to formulate this response in terms of a pseudo-current density, $\widetilde{\mathbf{J}}$, confined to the surface of the material, which relates to the curl of the magnetization,
\begin{equation}\label{eq.boundj}
    \mathbf{\nabla} \wedge \mathbf{M}=\widetilde{\textbf{J}}.
\end{equation}
The magnetic field strength resulting from a current source, $\mathbf{J_c}$, can be expressed using Amp\`ere's law
\begin{equation}\label{eq.current}
    \nabla \wedge \mathbf{H} = \mathbf{J_c}.
\end{equation}
Substituting \eqref{eq.boundj} and \eqref{eq.current} into the curl of \eqref{eq.mag}, noting $\mathbf{B}=\mathbf{\nabla} \wedge \mathbf{A}$ where $\mathbf{A}$ is the vector potential, results in the Poisson equation in free space,
\begin{equation}
    \nabla^2 \mathbf{A}=-\mu_0(\mathbf{J_c}+\widetilde{\textbf{J}}).
\end{equation}
As shown in many papers and textbooks~\cite{jackson}, the solution to the Poisson equation, for an arbitrary current density, $\mathbf{J}$, is given by
\begin{equation} \label{eq.vectorA}
    \mathbf{A\left(r\right)}=\mu_0 \int_{r'} \mathrm{d}^3\mathbf{r'}\ G(\mathbf{r},\mathbf{r'}) \mathbf{J(r')},
\end{equation}
where $G(\mathbf{r},\mathbf{r'})$ is the associated Green's function.

\begin{figure} [bht!]
\begin{center}
\begin{tikzpicture}
\draw [dashed] (-1.25,0) arc (180:360:1.25 and -0.5);
\draw [dashed] (-1.25,0) arc (180:360:1.25 and 0.5);
\draw (-1.25,0) -- (-1.25,-3.5);
\draw [dashed] (-1.25,-3.5) arc (180:360:1.25 and 0.5);
\draw [dashed] (-1.25,-3.5) arc (180:360:1.25 and -0.5);
\draw (1.25,-3.5) -- (1.25,0);  
\draw [->] (-3,-1.5) -- (3,-1.5);
\draw [<-] (0,1.5) -- (0,-4.5);
\draw [->] (-2,-3) -- (2,0);
\draw [<->] (0,0) -- (1.25,0);
\draw [dashed] (-1.25,0) -- (-2.5,0);
\draw [dashed] (-1.25,-3.5) -- (-2.5,-3.5);
\node at (3.25,-1.5) {$x$};
\node at (2.25,0) {$y$};
\node at (0,1.75) {$z$};
\node at (0.55,0.25) {$\rho_c$};
\node at (-3.2,0) {$z=L_1$};
\node at (-3.3,-3.5) {$z=L_2$};

\draw (0,0.5) ellipse (1.75 and 0.75);
\draw (-1.75,0.5) -- (-1.75,-4);
\draw (1.75,0.5) -- (1.75,-4);
\draw (-1.75,-4) arc (180:360:1.75 and 0.75);
\draw [dashed] (-1.75,-4) arc (180:360:1.75 and -0.75);
\draw [<->] (0,0.5) -- (1.75,0.5);
\draw [dashed] (1.75,0.5) -- (2.75,0.5);
\draw [dashed] (1.75,-4) -- (2.75,-4);
\node at (0.75,0.75) {$\rho_s$};
\node at (3.5,0.5) {$z=L_s/2$};
\node at (3.6,-4) {$z=-L_s/2$};
\end{tikzpicture}
\end{center}
\caption{Schematic diagram showing a high magnetic permeability cylinder of length, $L_s$, and radius, $\rho_s$, with planar end caps located at $z=\pm L_s/2$. This cylinder encloses an interior conducting open cylindrical shell of length, $L_c=L_1-L_2$, and radius, $\rho_c$.}
\label{magshield}
\end{figure}
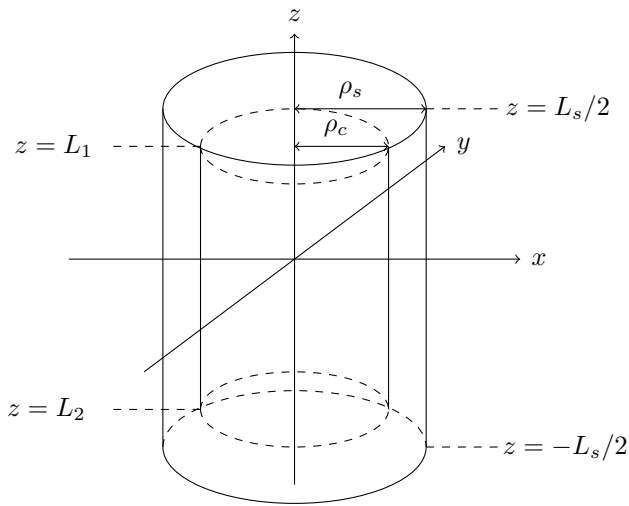

Let us now use the boundary condition on the magnetic field, \eqref{eq.btan}, to examine the effect of high-permeability material on the magnetic field generated by a specific current distribution. Consider a hollow high-permeability, $\mu_r\rightarrow \infty$, cylinder of radius, $\rho_s$, length, $L_s$, thickness, $d$, with high-permeability planar end caps located at $z=\pm L_s/2$ that is surrounded by free space (Fig.~\ref{magshield}). An arbitrary static current flows over an internal open coaxial cylinder of radius, $\rho_c<\rho_s$, and length, $L_1-L_2=L_c<L_s$, where $-L_s/2<L_2<L_1<L_s/2$, as shown in Fig.~\ref{magshield}.
Since high-permeability materials, such as mumetal, can have $\mu_r$ values >100,000 times that of air, the magnetic field must abruptly change direction at the boundary between its surface and air to satisfy the boundary conditions \eqref{eq.bperp}-\eqref{eq.btan}. When the shield is of sufficient thickness, the tangential components of the magnetic field at its boundary must be approximately zero. The boundary condition, \eqref{eq.btan}, on the interior surface of the exterior closed cylinder for the example depicted in Fig.~\ref{magshield}, requires that 
\begin{equation}
 B_\rho\bigg\rvert_{z=\pm L_s/2}\hspace{-30pt}\approx0, \quad\quad B_\phi\bigg\rvert_{z=\pm L_s/2,\rho=\rho_s}\hspace{-51pt}\approx0, \quad\hspace{35pt} B_z\bigg\rvert_{\rho=\rho_s}\approx0, \label{bbound}
\end{equation}
where the magnetic field, $\mathbf{B}=B_\rho~\boldsymbol{\hat{\rho}}+B_\phi~\boldsymbol{\hat{\phi}}+B_z~\mathbf{\hat{z}}$, is expressed in cylindrical polar coordinates. Previously, it has been found that the response of a high-permeability material deviates from that of a perfect magnetic conductor on the scale of $\mathcal{O}\left( \mu_r^{-1}\right)$~\cite{mu, mu1}. Therefore, we may assume for high-permeability materials, such as mumetal, the response can be approximated to that of a perfect magnetic conductor without introducing significant errors.\\

Due to the symmetries of the system it is natural to work in cylindrical coordinates. Following the formulation of Turner~\cite{turner}, we express the vector potential \eqref{eq.vectorA} due to an arbitrary current distribution, $\mathbf{J}=J_\rho(\mathbf{r'})~\boldsymbol{\hat{\rho}}+J_\phi(\mathbf{r'})~\boldsymbol{\hat{\phi}}+J_z(\mathbf{r'})~\hat{\mathbf{z}}$, as
\begin{equation}\label{eq.arho}
    A_{\rho}\left(\mathbf{r}\right)=\mu_0 \int_{r'} \mathrm{d}^3\mathbf{r'}\ G(\mathbf{r},\mathbf{r'})\left(J_{\rho}\left(\mathbf{r'}\right)\cos\left(\phi-\phi'\right)+J_{\phi}(\mathbf{r'})\sin\left(\phi-\phi'\right)\right),
\end{equation}
\begin{equation} \label{eq.aphi}
    A_{\phi}\left(\mathbf{r}\right)=-\mu_0 \int_{r'} \mathrm{d}^3\mathbf{r'}\ G(\mathbf{r},\mathbf{r'})\left(J_{\rho}\left(\mathbf{r'}\right)\sin\left(\phi-\phi'\right)-J_{\phi}(\mathbf{r'})\cos\left(\phi-\phi'\right)\right),
\end{equation}
\begin{equation} \label{eq.az}
    A_{z}\left(\mathbf{r}\right)=\mu_0 \int_{r'} \mathrm{d}^3\mathbf{r'}\ G(\mathbf{r},\mathbf{r'}) J_{z}(\mathbf{r'}).
\end{equation}
Since the current has been restricted to flow over an interior cylindrical shell centred radially about the origin, its radial components may be set to zero, resulting in the simplified vector potentials
\begin{equation}\label{eq.arho1}
    A_{\rho}\left(\mathbf{r}\right)=\mu_0 \int_{r'} \mathrm{d}^3\mathbf{r'}\ G(\mathbf{r},\mathbf{r'}) J_{\phi}(\mathbf{r'})\sin\left(\phi-\phi'\right),
\end{equation}
\begin{equation} \label{eq.aphi1}
    A_{\phi}\left(\mathbf{r}\right)=\mu_0 \int_{r'} \mathrm{d}^3\mathbf{r'}\ G(\mathbf{r},\mathbf{r'}) J_{\phi}(\mathbf{r'})\cos\left(\phi-\phi'\right).
\end{equation}
Substituting the Green's function solution from \eqref{eq.vectorA} in cylindrical coordinates~\cite{jackson},
\begin{equation}\label{green}
    G(\mathbf{r,r'}) =\frac{1}{4\pi^2}\sum_{m=-\infty}^{\infty}e^{im(\phi-\phi')}\int_{-\infty}^{\infty}\mathrm{d}k \ e^{ik\left(z-z'\right)}
     I_m(|k|\rho_<)K_m(|k|\rho_>),
\end{equation}
into (\ref{eq.az})-(\ref{eq.aphi1}), the components of the vector potential may be expressed as
\begin{align}\nonumber
    A_{\rho}\left(\rho,\phi,z\right)=-\frac{i\mu_0\rho'}{4\pi}\sum_{m=-\infty}^{\infty}\int_{-\infty}^{\infty}\mathrm{d}k \ e^{im\phi}e^{ikz}
     \Big[I_{m-1}(|k|\rho_<)K_{m-1}(|k|\rho_>) \qquad\qquad\qquad\qquad\qquad\qquad \\ \label{eq.arn} -I_{m+1}(|k|\rho_<)K_{m+1}(|k|\rho_>)\Big]J_\phi^m(k),
\end{align}
\begin{align}\nonumber
    A_{\phi}\left(\rho,\phi,z\right)=\frac{\mu_0\rho'}{4\pi}\sum_{m=-\infty}^{\infty}\int_{-\infty}^{\infty}\mathrm{d}k \ e^{im\phi}e^{ikz}
     \Big[I_{m-1}(|k|\rho_<)K_{m-1}(|k|\rho_>) \ \ \ \ \qquad\qquad\qquad\qquad\qquad\qquad \\ \label{eq.apn} +I_{m+1}(|k|\rho_<)K_{m+1}(|k|\rho_>)\Big]J_\phi^m(k),
\end{align}
\begin{align} \label{eq.azn}
    A_{z}\left(\rho,\phi,z\right)=\frac{\mu_0\rho'}{2\pi}\sum_{m=-\infty}^{\infty}\int_{-\infty}^{\infty}\mathrm{d}k \ e^{im\phi}e^{ikz}
     I_m(|k|\rho_<)K_{m}(|k|\rho_>)J_z^m(k),
\end{align}
where $\rho'$ is the radius of the cylinder and $\rho_{>}$, $\rho_{<}$ is either $\rho$, $\rho'$ if $\rho>\rho'$ or $\rho'$, $\rho$ if $\rho<\rho'$, respectively. Equations \eqref{eq.arn}-\eqref{eq.azn} contain Fourier transforms of the current densities,
\begin{equation}\label{eq.fp1}
    J_\phi^m(k)=\frac{1}{2\pi}\int_{0}^{2\pi}\mathrm{d}\phi' \ e^{-im\phi'}\int_{-\infty}^{\infty}\mathrm{d}z'\ e^{-ikz'}J_\phi\left(\phi',z'\right), 
\end{equation}
\begin{equation}\label{eq.fz1}
    J_z^m(k)=\frac{1}{2\pi}\int_{0}^{2\pi}\mathrm{d}\phi' \ e^{-im\phi'}\int_{-\infty}^{\infty}\mathrm{d}z'\ e^{-ikz'}J_z\left(\phi',z'\right),
\end{equation}
with their corresponding inverse transforms given by
\begin{equation}\label{eq.ifp}
    J_\phi(\phi',z')=\frac{1}{2\pi}\sum_{m=-\infty}^{\infty} \int_{-\infty}^{\infty}\mathrm{d}k\ e^{im\phi'}e^{ikz'}J_\phi^m\left(k\right),
\end{equation}
\begin{equation} \label{eq.ifz}
    J_z(\phi',z')=\frac{1}{2\pi}\sum_{m=-\infty}^{\infty} \int_{-\infty}^{\infty}\mathrm{d}k\ e^{im\phi'}e^{ikz'}J_z^m\left(k\right). 
\end{equation}

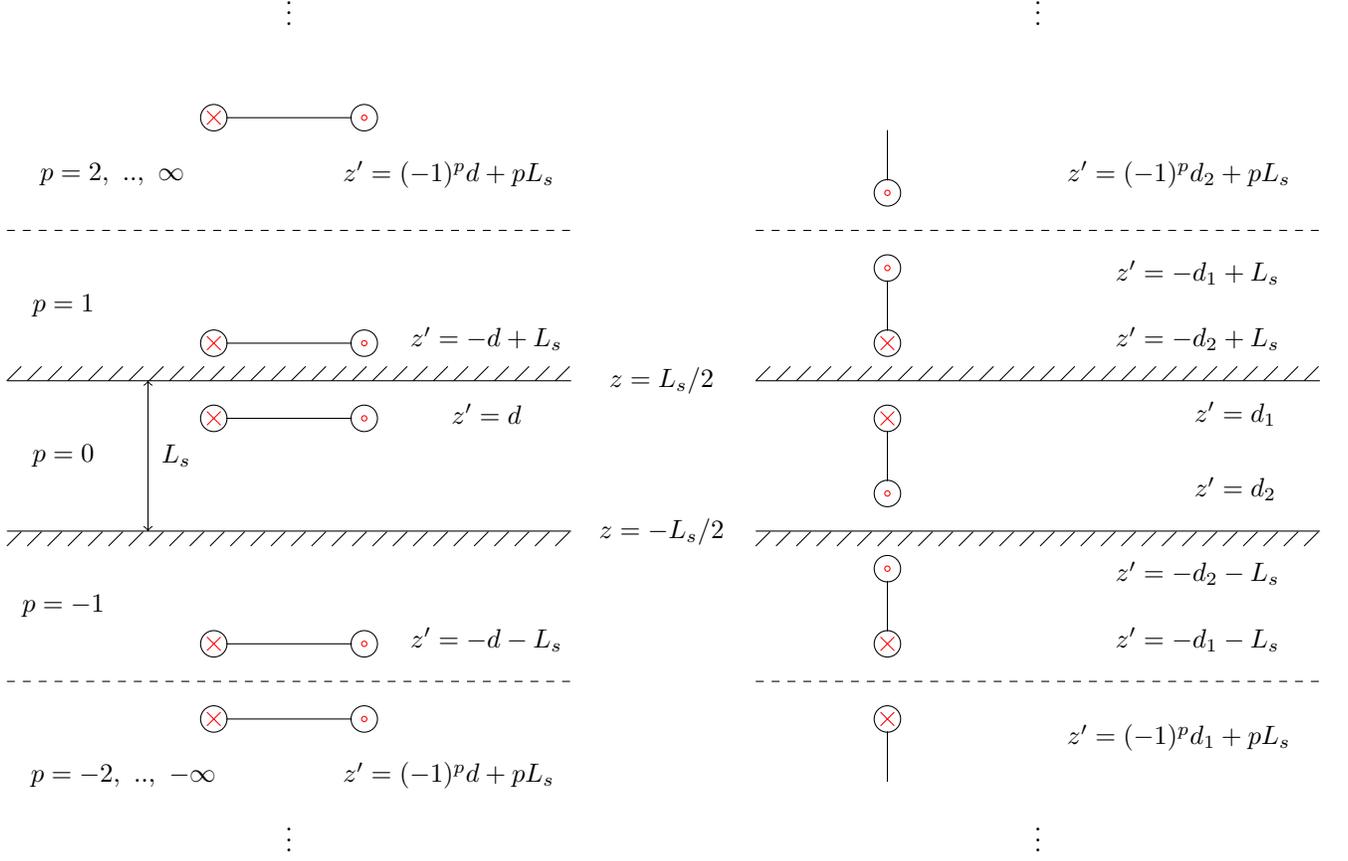
\begin{figure*}[!htb]
    \centering
    \hspace{0pt}
    {\begin{tikzpicture}[scale=0.5]
    
    \draw (-2.5,2) -- (12.5,2);
    \draw (-2.5,-2) -- (12.5,-2);
    \draw [dashed] (-2.5,6) -- (12.5,6);
    \draw [dashed] (-2.5,-6) -- (12.5,-6);

    \node at (5, 12) {$\vdots$};
    \node at (5, -10) {$\vdots$};
    
    \draw[decorate,decoration={border,amplitude=0.27cm,segment length=0.27cm}] (-2.5,2) - - (13,2); 
    \draw[decorate,decoration={border,amplitude=0.27cm,segment length=0.27cm}] (-2.5,-2-0.1768*2.25) - - (13,-2-0.1768*2.25); 
    
    \draw (3,1) circle (10pt);
    \draw (3,1) node[cross,red] {};
    \draw (7,1) circle (10pt);
    \draw (7,1) [red] circle (2pt);
    \draw (3.33,1) -- (6.67,1);
    \node at (10.25,1.125) {$z'=d$};
    \node at (-1,0) {$p=0$};
    
    \draw (3,3) circle (10pt);
    \draw (3,3) node[cross,red] {};
    \draw (7,3) circle (10pt);
    \draw (7,3) [red] circle (2pt);
    \draw (3.33,3) -- (6.67,3);
    \node at (10.25,3.125) {$z'=-d+L_s$};
    \node at (-1,4) {$p=1$};
    
    \draw (3,9) circle (10pt);
    \draw (3,9) node[cross,red] {};
    \draw (7,9) circle (10pt);
    \draw (7,9) [red] circle (2pt);
    \draw (3.33,9) -- (6.67,9);
    \node at (9.25,7.5) {$z'=(-1)^pd+pL_s$};
    \node at (0.3,7.5) {$p=2, \ .., \ \infty$};
    
    \draw (3,-5) circle (10pt);
    \draw (3,-5) node[cross,red] {};
    \draw (7,-5) circle (10pt);
    \draw (7,-5) [red] circle (2pt);
    \draw (3.33,-5) -- (6.67,-5);
    \node at (10.25,-4.875) {$z'=-d-L_s$};
    \node at (-1,-4) {$p=-1$};
    
    \draw (3,-7) circle (10pt);
    \draw (3,-7) node[cross,red] {};
    \draw (7,-7) circle (10pt);
    \draw (7,-7) [red] circle (2pt);
    \draw (3.33,-7) -- (6.67,-7);
    \node at (9.25,-8.5) {$z'=(-1)^pd+pL_s$};
    \node at (0.6,-8.5) {$p=-2, \ .., \ -\infty$};
    
    \draw [<->] (1.25,2) -- (1.25,-2);
    \node at (2,0) {$L_s$};

    \end{tikzpicture}}%
    {\begin{tikzpicture}[scale=0.5]
    
    \draw (-2.5,2) -- (12.5,2);
    \draw (-2.5,-2) -- (12.5,-2);
    \draw [dashed] (-2.5,6) -- (12.5,6);
    \draw [dashed] (-2.5,-6) -- (12.5,-6);
    
    \node at (5, 12) {$\vdots$};
    \node at (5, -10) {$\vdots$};
    
    \draw[decorate,decoration={border,amplitude=0.27cm,segment length=0.27cm}] (-2.5,2) - - (13,2); 
    \draw[decorate,decoration={border,amplitude=0.27cm,segment length=0.27cm}] (-2.5,-2-0.1768*2.25) - - (13,-2-0.1768*2.25); 
    
    \draw (1,1) circle (10pt);
    \draw (1,1) node[cross,red] {};
    \draw (1,-1) circle (10pt);
    \draw (1,-1) [red] circle (2pt);
    \draw (1,0.67) -- (1,-0.67);
    \node at (10.25,1.125) {$z'=d_1$};
    \node at (10.25,-0.875) {$z'=d_2$};
    
    \draw (1,3) circle (10pt);
    \draw (1,3) node[cross,red] {};
    \draw (1,5) circle (10pt);
    \draw (1,5) [red] circle (2pt);
    \draw (1,3.33) -- (1,4.67);
    \node at (9.25,3.125) {$z'=-d_2+L_s$};
    \node at (9.25,4.875) {$z'=-d_1+L_s$};
    
    \draw (1,7) circle (10pt);
    \draw (1,7) [red] circle (2pt);
    \draw (1,7.33) -- (1,8.67);
    \node at (8.75,7.5) {$z'=(-1)^pd_2+pL_s$};
    
    \draw (1,-5) circle (10pt);
    \draw (1,-5) node[cross,red] {};
    \draw (1,-3) circle (10pt);
    \draw (1,-3) [red] circle (2pt);
    \draw (1,-3.33) -- (1,-4.67);
    \node at (9.25,-3.125) {$z'=-d_2-L_s$};
    \node at (9.25,-4.875) {$z'=-d_1-L_s$};
    
    \draw (1,-7) circle (10pt);
    \draw (1,-7) node[cross,red] {};
    \draw (1,-7.33) -- (1,-8.67);
    \node at (8.75,-7.5) {$z'=(-1)^pd_1+pL_s$};
    
    \node at (-5,2) {$z=L_s/2$};
    \node at (-5,-2) {$z=-L_s/2$};

    \end{tikzpicture}}%
    \caption{Schematic diagram showing how azimuthal (left) and axial (right) currents are reflected by two infinite parallel planes (hatched) of infinite extension and permeability (i.e. perfect magnetic conductors) located at $z=\pm L_s/2$. (left) Azimuthal currents located in the $z'=d$ plane are translated, by the reflection process, to the $z'=(-1)^pd+pL_s$ planes. (right) Axial currents flowing from $z'=d_1$ to $z'=d_2$ are reflected to positions $z'=(-1)^pd_1+pL_s$ and $z'=(-1)^pd_2+pL_s$, respectively, with odd reflections reversing the initial current flow direction. Black circles show cross-sections of wires carrying current into (red crosses) and out of (red circles) the page. }
    \label{fig.reflect}
\end{figure*}

As a result of this formulation, we can now combine methods for matching the boundary conditions at the radial interface, akin to the formulation of passive screening of magnetic field gradients for MRI \cite{ap}, with the method of mirror images, accounting for the effect of the planar end caps, to determine the effect of the high-permeability material on the overall magnetic field profile. Due to the cylindrical symmetry of the system, the radial boundary condition may be satisfied by matching the azimuthal Fourier modes in the magnetic field, generated by the cylindrical current source, through the use of a pseudo-current density induced on an infinite cylinder. As the planar end caps are spatially orthogonal to the pseudo-current generated by the infinite cylinder, any image current formed by applying the method of mirror images continues to satisfy the radial boundary condition. Consequently, we can decouple the boundary conditions on the high-permeability cylinder and at the planar end cap boundaries. This must be done sequentially to generate mirror images of the pseudo-current density induced on the high-permeability cylindrical shell and, hence, satisfy the boundary conditions over the entire domain of the closed finite high-permeability cylinder. As a result, using \eqref{eq.arn}-\eqref{eq.azn} and the usual formulation of the method of mirror images with two infinite parallel planes, depicted in Fig.~\ref{fig.reflect}, the vector potential in the region $\rho_c\leq\rho\leq\rho_s$, including the effect of the high-permeability cylinder, may be written in terms of an infinite summation over the reflected image currents, 
\begin{align}\nonumber
    A_{\rho}\left(\rho,\phi,z\right)=-\frac{i\mu_0}{4\pi}\sum_{m=-\infty}^{\infty}\sum_{p=-\infty}^{\infty}\int_{-\infty}^{\infty}\mathrm{d}k \ e^{im\phi}e^{ikz}
     \Bigg\{\rho_c\Big[I_{m-1}(|k|\rho_c)K_{m-1}(|k|\rho) \qquad\qquad\qquad\qquad\qquad
     \\  -I_{m+1}(|k|\rho_c)K_{m+1}(|k|\rho)\Big]J_\phi^{mp}(k)+\rho_s\Big[I_{m-1}(|k|\rho)K_{m-1}(|k|\rho_s) \nonumber \qquad\qquad\\ \label{eq.arnew} -I_{m+1}(|k|\rho)K_{m+1}(|k|\rho_s)\Big]\widetilde{J_\phi^{mp}}(k)\Bigg\},
\end{align}
\begin{align}\nonumber
    A_{\phi}\left(\rho,\phi,z\right)=\frac{\mu_0}{4\pi}\sum_{m=-\infty}^{\infty}\sum_{p=-\infty}^{\infty}\int_{-\infty}^{\infty}\mathrm{d}k \ e^{im\phi}e^{ikz}
     \Bigg\{\rho_c\Big[I_{m-1}(|k|\rho_c)K_{m-1}(|k|\rho) \qquad\qquad\qquad\qquad\qquad \ \ \ \
     \\  +I_{m+1}(|k|\rho_c)K_{m+1}(|k|\rho)\Big]J_\phi^{mp}(k)+\rho_s\Big[I_{m-1}(|k|\rho)K_{m-1}(|k|\rho_s) \nonumber\qquad\qquad \\ \label{eq.apnew} +I_{m+1}(|k|\rho)K_{m+1}(|k|\rho_s)\Big]\widetilde{J_\phi^{mp}}(k)\Bigg\},
\end{align}
\begin{align} 
    A_{z}\left(\rho,\phi,z\right)=\frac{\mu_0}{2\pi}\sum_{m=-\infty}^{\infty}\sum_{p=-\infty}^{\infty}\int_{-\infty}^{\infty}\mathrm{d}k \ e^{im\phi}e^{ikz}
     \Big[\rho_cI_{m}(|k|\rho_c)K_{m}(|k|\rho)J_z^{mp}(k) \qquad\qquad\qquad\qquad\qquad \ \ \ \ \nonumber \\ \label{eq.aznew} +\rho_sI_{m}(|k|\rho)K_{m}(|k|\rho_s)\widetilde{J_z^{mp}}(k)\Big],
\end{align}
where
\begin{equation}\label{eq.fp}
    J_\phi^{mp}(k)=\frac{1}{2\pi}\int_{0}^{2\pi}\mathrm{d}\phi' \ e^{-im\phi'}\int_{-\infty}^{\infty}\mathrm{d}z'\ e^{-ikz'}J_\phi^p\left(\phi',z'\right), 
\end{equation}
\begin{equation}\label{eq.fz}
    J_z^{mp}(k)=\frac{1}{2\pi}\int_{0}^{2\pi}\mathrm{d}\phi' \ e^{-im\phi'}\int_{-\infty}^{\infty}\mathrm{d}z'\ e^{-ikz'}J_z^p\left(\phi',z'\right),
\end{equation}
are the Fourier transforms of the $p^{th}$ reflected image current and $\widetilde{J^{mp}_{\phi,z}}(k)$ is the Fourier transform of the $p^{th}$ reflected pseudo-current induced on the high-permeability cylinder. Fig.~\ref{fig.reflect} depicts how azimuthal, $J_\phi^p\left(\phi',z'\right)$, and axial, $J_z^p\left(\phi',z'\right)$,  image currents are generated by two parallel planar perfect magnetic conductors.  

Writing the magnetic field in cylindrical coordinates,
\begin{equation} \label{curl}
    \mathbf{B}=\nabla \wedge \mathbf{A} = \left(\frac{1}{\rho}\frac{\partial A_z}{\partial \phi}- \frac{\partial A_\phi}{\partial z}\right)\boldsymbol{\hat{\rho}}+\left(\frac{\partial A_\rho}{\partial z}-\frac{\partial A_z}{\partial \rho}\right)\boldsymbol{\hat{\phi}}+\frac{1}{\rho}\left(\frac{\partial}{\partial \rho}(\rho A_\phi)-\frac{\partial A_\rho}{\partial \phi}\right)\mathbf{\hat{z}},
\end{equation}
the boundary conditions, (\ref{bbound}), on the magnetic field at $\rho=\rho_s$ are
\begin{align}\label{eq.condb1}
    \frac{1}{\rho}\left(\frac{\partial}{\partial \rho}(\rho A_\phi)-\frac{\partial A_\rho}{\partial \phi}\right)\Bigg\rvert_{\rho=\rho_s} =0,
\end{align}
\begin{align}\label{eq.condb2}
    \left(\frac{\partial A_\rho}{\partial z}-\frac{\partial A_z}{\partial \rho}\right)\Bigg\rvert_{\rho=\rho_s} =0.
\end{align}
Substituting \eqref{eq.arnew}-\eqref{eq.aznew} into \eqref{eq.condb1}-\eqref{eq.condb2}, the Fourier transformed pseudo-current density on the cylindrical wall of the high-permeability cylinder is found to be
\begin{equation}\label{eq.pseudo}
    \widetilde{J_{\phi,z}^{mp}}(k)=-\frac{\rho_cI'_{m}(|k|\rho_c)K_{m}(|k|\rho_s)}{\rho_sI_{m}(|k|\rho_s)K'_{m}(|k|\rho_s)}J_{\phi,z}^{mp}(k),
\end{equation}
where $I'_m(z)$ and $K'_m(z)$ are the derivatives with respect to $z$ of $I_m(z)$ and $K_m(z)$, respectively. Using the continuity equation, \eqref{eq.ifp}-\eqref{eq.aznew}, and \eqref{eq.pseudo}, the magnetic field interior to the conducting cylinder, for all points $\rho<\rho_c$, is given by
\begin{align}\label{eq.Brc}
    B_{\rho}\left(\rho,\phi,z\right)=\frac{i\mu_0\rho_c}{2\pi}\sum_{m=-\infty}^{\infty}\sum_{p=-\infty}^{\infty}\int_{-\infty}^{\infty}\mathrm{d}k \ ke^{im\phi}e^{ikz}
     I'_{m}(|k|\rho)R_m(k,\rho_c,\rho_s)J_\phi^{mp}(k),
\end{align}
\begin{align}\label{eq.Bpc}
    B_{\phi}\left(\rho,\phi,z\right)=-\frac{\mu_0\rho_c}{2\pi\rho}\sum_{m=-\infty}^{\infty}\sum_{p=-\infty}^{\infty}\int_{-\infty}^{\infty}\mathrm{d}k \ m\frac{|k|}{k}e^{im\phi}e^{ikz}
     I_{m}(|k|\rho) R_m(k,\rho_c,\rho_s)J_\phi^{mp}(k),
\end{align}
\begin{align}\label{eq.Bzc}
    B_{z}\left(\rho,\phi,z\right)=-\frac{\mu_0\rho_c}{2\pi}\sum_{m=-\infty}^{\infty}\sum_{p=-\infty}^{\infty}\int_{-\infty}^{\infty}\mathrm{d}k \ |k|e^{im\phi}e^{ikz}
     I_{m}(|k|\rho)R_m(k,\rho_c,\rho_s)J_\phi^{mp}(k),
\end{align}
where $R_m(k,\rho_c,\rho_s)=K'_{m}(|k|\rho_c)-\frac{I'_{m}(|k|\rho_c)K_{m}(|k|\rho_s)}{I_{m}(|k|\rho_s)}$.
In order to determine the magnetic field generated by an arbitrary cylindrical current source, we must construct an orthogonal basis defined on a finite cylinder, which accounts for the mirror images. To do this we use a modified Fourier basis, defining the $p^{\textnormal{th}}$ reflected azimuthal current to be
\begin{align}
     J_{\phi}^p(\phi',z')&=\left(T^p_e(z';L_1,L_2,L_s)+T^p_o(z';L_1,L_2,L_s)\right)\Bigg[\sum_{n=1}^N W_{n0}\sin\left(\frac{n\pi\left((-1)^p\left(z'-pL_s\right)-L_2\right)}{L_c}\right)\nonumber \\&\qquad\qquad+\sum_{n=1}^N\sum_{m=1}^M\left(W_{nm}\cos(m\phi')+Q_{nm}\sin(m\phi')\right)\cos\left(\frac{n\pi\left((-1)^p\left(z'-pL_s\right)-L_2\right)}{L_c}\right)\Bigg] \label{eq.jphinew},
\end{align}
in which
\begin{equation}
    T^p_e(z';L_1,L_2,L_s)=\left(H\left(z'-L_2-pL_s\right)-H\left(z'-L_1-pL_s\right)\right)\left(\frac{1+(-1)^{p}}{2}\right),
\end{equation}
\begin{equation}
    T^p_o(z';L_1,L_2,L_s)=\left(H\left(z'+L_1-pL_s\right)-H\left(z'+L_2-pL_s\right)\right)\left(\frac{1-(-1)^p}{2}\right),
\end{equation}
where $H(x)$ is the Heaviside function and $\left(W_{n0},W_{nm},Q_{nm}\right)$ are Fourier coefficients to be determined. Substituting \eqref{eq.jphinew} into \eqref{eq.Brc}-\eqref{eq.Bzc}, we derive a set of governing equations which relate the magnetic field to the set of weighted Fourier coefficients,
\begin{equation}\label{eq.brf}
    B_{\rho}\left(\rho,\phi,z\right)= \sum_{n=1}^NW_{n0} F_{n}\left(\rho,z\right)+\sum_{n=1}^N\sum_{m=1}^M\left(W_{nm}G^w_{nm}\left(\rho,\phi,z\right)+Q_{nm}G^q_{nm}\left(\rho,\phi,z\right)\right),
\end{equation}
\begin{equation}\label{eq.bpf}
    B_{\phi}\left(\rho,\phi,z\right)=\sum_{n=1}^N\sum_{m=1}^M\left(W_{nm}H^w_{nm}\left(\rho,\phi,z\right)+Q_{nm}H^q_{nm}\left(\rho,\phi,z\right)\right),
\end{equation}
\begin{equation}\label{eq.bzf}
    B_{z}\left(\rho,\phi,z\right)=\sum_{n=1}^NW_{n0} D_{n}\left(\rho,z\right)+\sum_{n=1}^N\sum_{m=1}^M\left(W_{nm}S^w_{nm}\left(\rho,\phi,z\right)+Q_{nm}S^q_{nm}\left(\rho,\phi,z\right)\right),
\end{equation}
where the functions $F_{n}\left(\rho,z\right)$, $G^{w,q}_{nm}\left(\rho,\phi,z\right)$, $H^{w,q}_{nm}\left(\rho,\phi,z\right)$, $D_{n}\left(\rho,z\right)$, and $S^{w,q}_{nm}\left(\rho,\phi,z\right)$ are defined in Appendix~A. Having determined the magnetic field produced by an arbitrary cylindrical current, we may now use an inverse method to solve the system of governing equations, \eqref{eq.brf}-\eqref{eq.bzf}, to determine the unknown Fourier coefficients $\left(W_{n0},W_{nm},Q_{nm}\right)$ for a specified target magnetic field. Following work done by Carlson \emph{et al.}~\cite{doi:10.1002/mrm.1910260202}, this may be done by a least squares minimization with the addition of a penalty term to regularize the problem. This regularization term may take many forms, with individual contributions to it representing, for example, the curvature of a given wire geometry, the power consumption, or any other physical parameter that depends quadratically on the geometry of the coil. In this work we focus on the overall power dissipated in the cylindrical current flow, but this choice is somewhat arbitrary since all of the regularization parameters act to achieve the same general goal. If the regularization term is large, the result is a well-conditioned inverse problem that yields a simple current flow, but reduced field fidelity. On the other hand, if the regularization term is small, then the result is a less well conditioned inverse problem that yields a more intricate pattern of current flow but a higher-fidelity magnetic field. The power dissipation in the conducting cylinder of thickness, $t$, and resistivity, $\varrho$, is given by
\begin{equation} \label{eq.power}
    P=\frac{\rho_c\varrho}{t}\int_{L_2}^{L_1}\mathrm{d}z'\int_{0}^{2\pi}\mathrm{d}\phi'\ |J_z(\phi',z')|^2+|J_\phi(\phi',z')|^2,
\end{equation}
which, when integrated over the surface of the cylinder, using the continuity equation and \eqref{eq.jphinew}, gives
\begin{equation}
    P=\frac{\rho_c\varrho}{t}\left[\sum_{n=1}^NW_{n0}^2\pi L_c+\sum_{n=1}^N\sum_{m=1}^M\left(W_{nm}^2+Q_{nm}^2\right)\left(\frac{\pi L_c}{2}+\frac{m^2L_c^3}{2\pi n^2 \rho_c^2}\right)\right].
\end{equation}

We now construct a cost function using a least squares optimization procedure,
\begin{equation} \label{eq.functional}
\Phi=\sum_k \left[\mathbf{B}^{\mathrm{desired}}\left(\mathbf{r}_k\right)-\mathbf{B}\left(\mathbf{r}_k\right)\right]^2 + \beta P,
\end{equation}
evaluated at $K$ target field points, where $\beta$ is a weighting parameter chosen such that the physical parameters may be adjusted to achieve specified physical constraints. The minimization is achieved by taking the derivative of the cost function with respect to the Fourier coefficients,
\begin{equation}\label{eq.mindif}
    \frac{\partial \Phi}{\partial W_{i0}}=0, \qquad \frac{\partial \Phi}{\partial W_{ij}}=0, \qquad \frac{\partial \Phi}{\partial Q_{ij}}=0, \qquad i,j>1,
\end{equation}
allowing the optimal Fourier coefficients to be found by matrix inversion for any physically attainable target magnetic field. The current density is then related, through the continuity equation, to the streamfunction
\begin{align}\nonumber
    \varphi(\phi',z')=\left(H\left(z'-L_1\right)-H\left(z'-L_2\right)\right)\Bigg[\sum_{n=1}^N \frac{L_c}{n\pi}W_{n0}\cos\left(\frac{n\pi\left(z'-L_2\right)}{L_c}\right)\qquad\qquad\qquad\qquad\qquad\qquad \\  -\sum_{n=1}^N\sum_{m=1}^M\frac{L_c}{n\pi}\left(W_{nm}\cos(m\phi')+Q_{nm}\sin(m\phi')\right)\sin\left(\frac{n\pi\left(z'-L_2\right)}{L_c}\right)\Bigg],
\end{align}
on the inner conducting cylindrical surface.\\ 

Ideally, the Fourier series would have an infinite number of coefficients. However, truncating the series at a large finite number of terms still provides accurate solutions for sufficiently regularized problems. This is because higher-order terms contribute less at distances much larger than their spatial period. When designing fields in regions close to the conducting surface, if a sufficiently large number of coefficients is chosen, the real-world field fidelity is limited by the accuracy to which the current continuum can be approximated by discretized wire configurations, rather than by the number of Fourier coefficients. Typically, setting N=200 and equating M to the degree of the desired field harmonic is adequate. An approximate solution to the current continuum is then found by discretizing the streamfunction into $N_{\varphi}$ values, where $\varphi_j=\textnormal{min}\ \varphi+(j-1/2)\Delta\varphi, j=1,..., N_{\varphi}$, separated by $\Delta \varphi = \frac{\textnormal{max}\ \varphi-\textnormal{min}\ \varphi}{N_{\varphi}}$. The number of contours should be maximized according to the constraints in manufacturing because the accuracy of the model depends on the approximation of the continuum. For situations where only coarse discretization is possible, we recommend that a form of discrete optimization is used instead. When physically manufacturing these structures, the distance of the wires from the high-permeability material is important. Without the passive shield, one can determine how accurately the discrete coil represents the current continuum by using the elemental Biot–Savart law to calculate the error as $N_{\varphi}$ is adjusted.

\section{Results}
We now analyze our theoretical model by designing and testing hybrid active--passive magnetic field-generating systems. Regarding the validation of our calculations, we first note that, as expected from previous work~\cite{Solenoid1, solenoid2, doi:10.1063/1.1719514} and shown in the Supplementary Material, our calculations confirm that the optimal coil design for generating a constant axial field inside a closed cylindrical perfect magnetic shield is a perfect solenoid that runs along the full length of the cylindrical shield.\\

\begin{figure*}[!htb]
     \centering
         {\includegraphics[scale=0.3185]{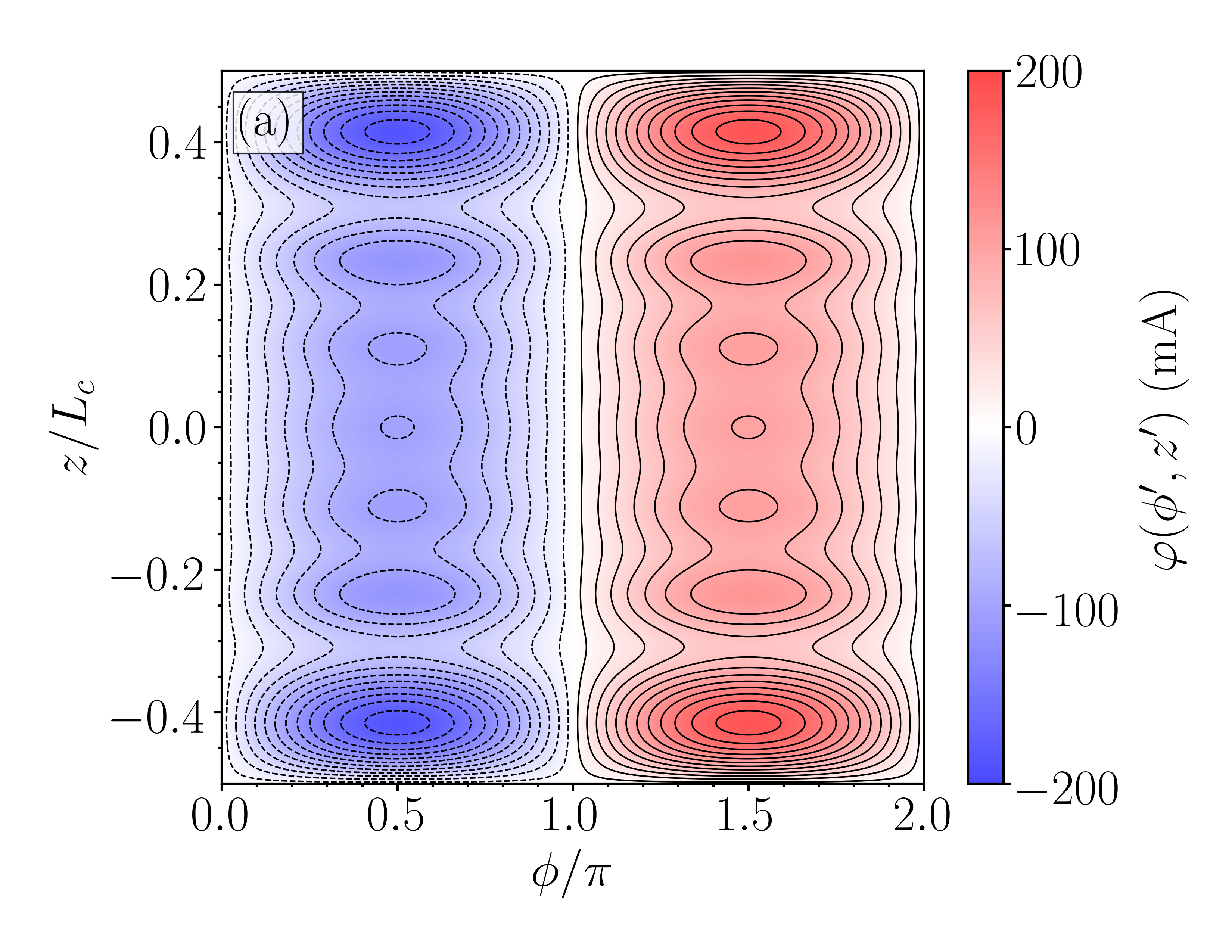}}
         {\includegraphics[scale=0.3185]{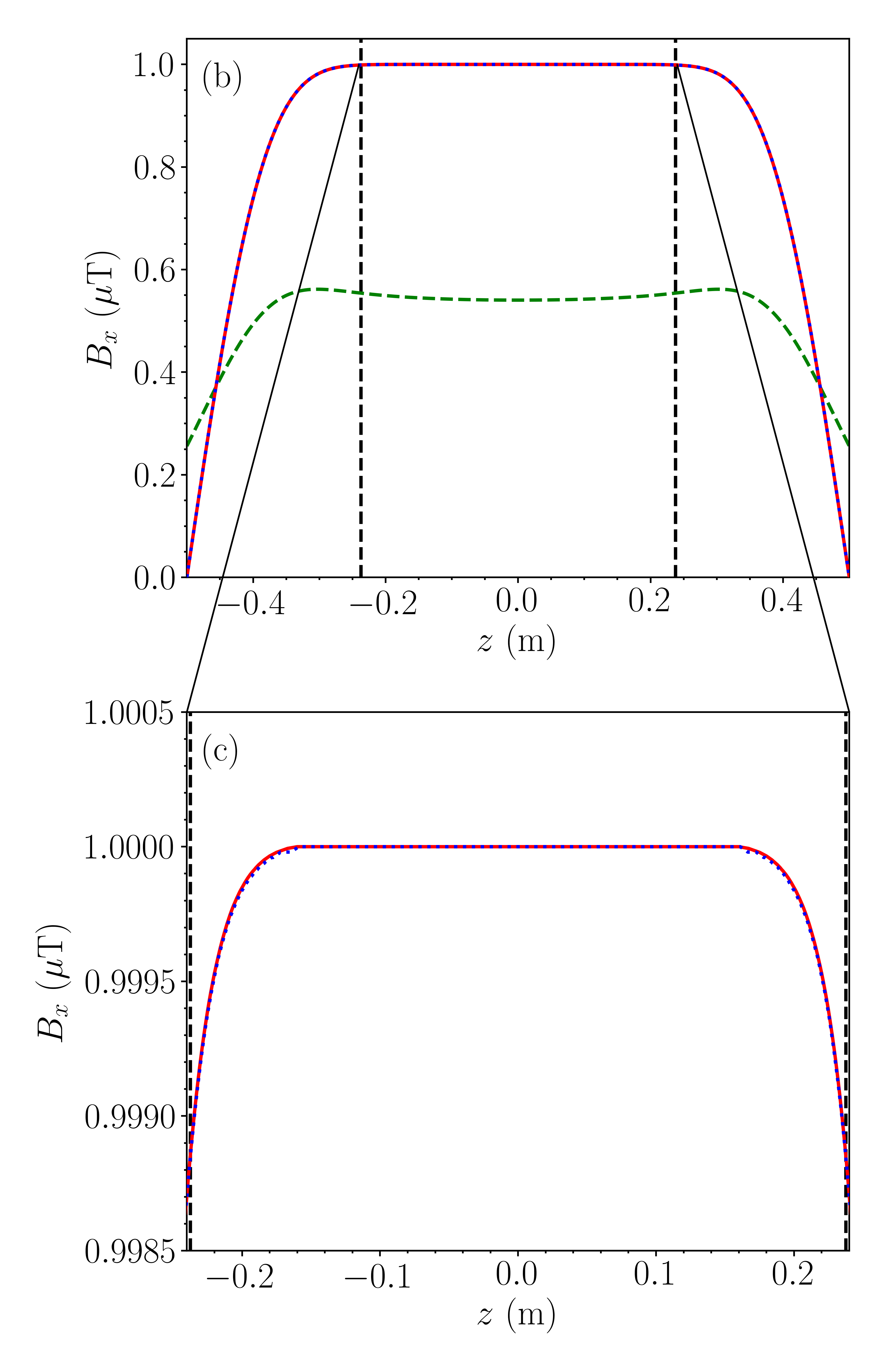}}
        \caption{Wire layouts (a) and performance (b--c) of an optimized hybrid active--passive constant transverse field-generating system in which current flows on a cylinder of length $L_c=0.95$ m and radius $\rho_c=0.245$ m. The wire layouts are optimized to generate a constant transverse field, $B_x=1$ $\mu$T, across the cylinder and normal to its axis of symmetry. The current-carrying cylinder is placed symmetrically inside a perfect closed magnetic shield of length $L_s=1$ m and radius $\rho_s=0.25$ m and the magnetic field is optimized between $\rho=[0,\rho_c/2]$ and $z=\pm{L_c}/4$; dashed black lines in (b--c). The least squares optimization was performed with parameters $N=200$, $M=1$, and $\beta=5.95\times10^{-12}$. (a) Color map of the optimal current streamfunction on the cylinder [blue and red shaded regions correspond to the flow of current in opposite senses respectively and their intensity shows the streamfunction magnitude from low (white) to high (intense color)]. Solid and dashed black curves represent discrete wires with opposite senses of current flow, approximating the current continuum with $N_\varphi=24$ contour levels. (b) Transverse magnetic field, $B_x$, versus axial position, $z$, calculated from the current continuum in (a) in three ways: analytically using \eqref{eq.brf}-\eqref{eq.bzf} (solid red curve); numerically using COMSOL Multiphysics\textsuperscript{\textregistered} Version 5.3a and modelling the high-permeability cylinder as a perfect magnetic conductor (blue dotted curve); numerically \emph{without} the high-permeability cylinder and using the Biot–Savart law with $N_{\varphi}=100$ contour levels (dashed green). (c) Enlarged section of (b) emphasizing the high level of field uniformity and the agreement between the numerical and analytical results over the optimization region.}
        \label{fig.bxnew}
\end{figure*}

\begin{figure*}[!htb]
     \centering
         {\includegraphics[scale=0.3185]{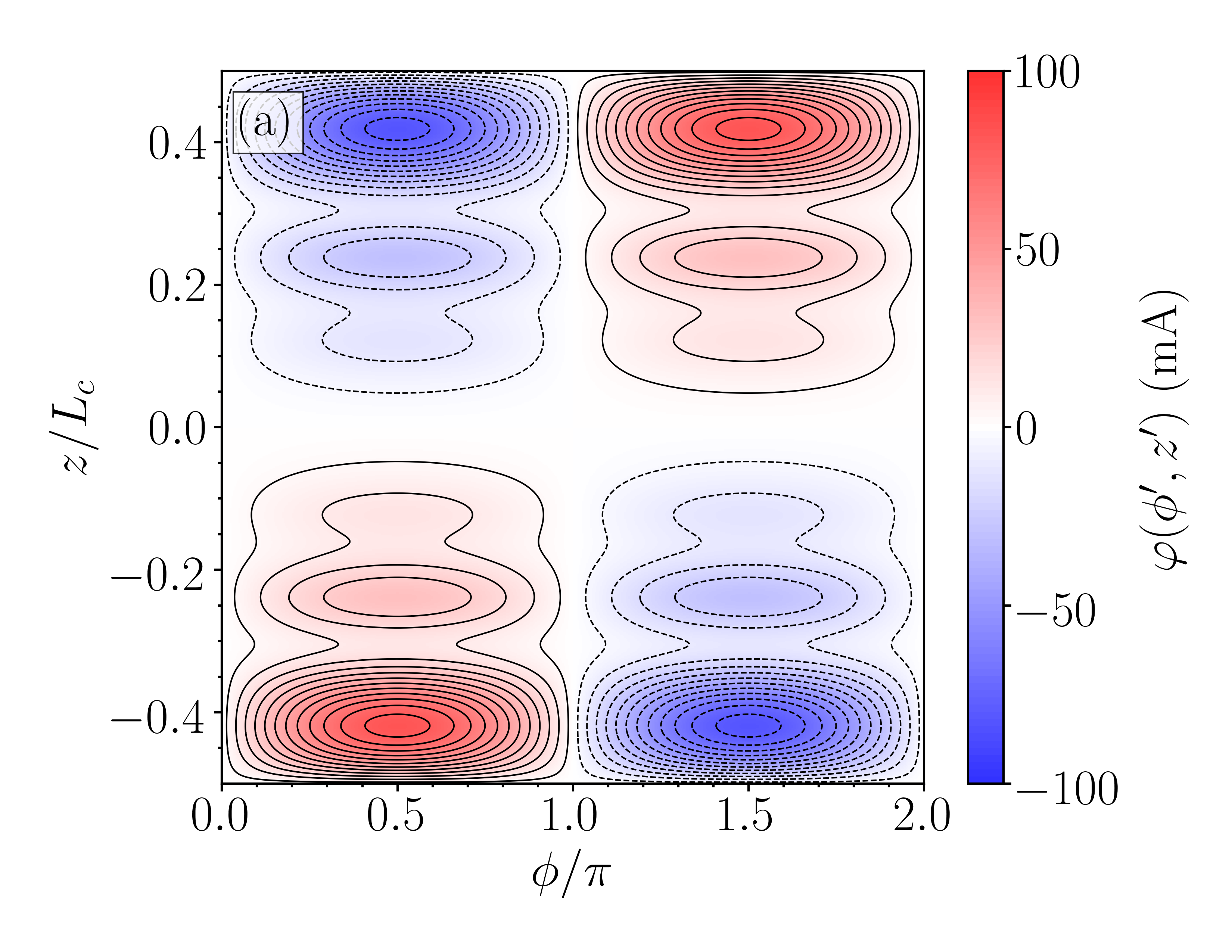}}
         {\includegraphics[scale=0.3185]{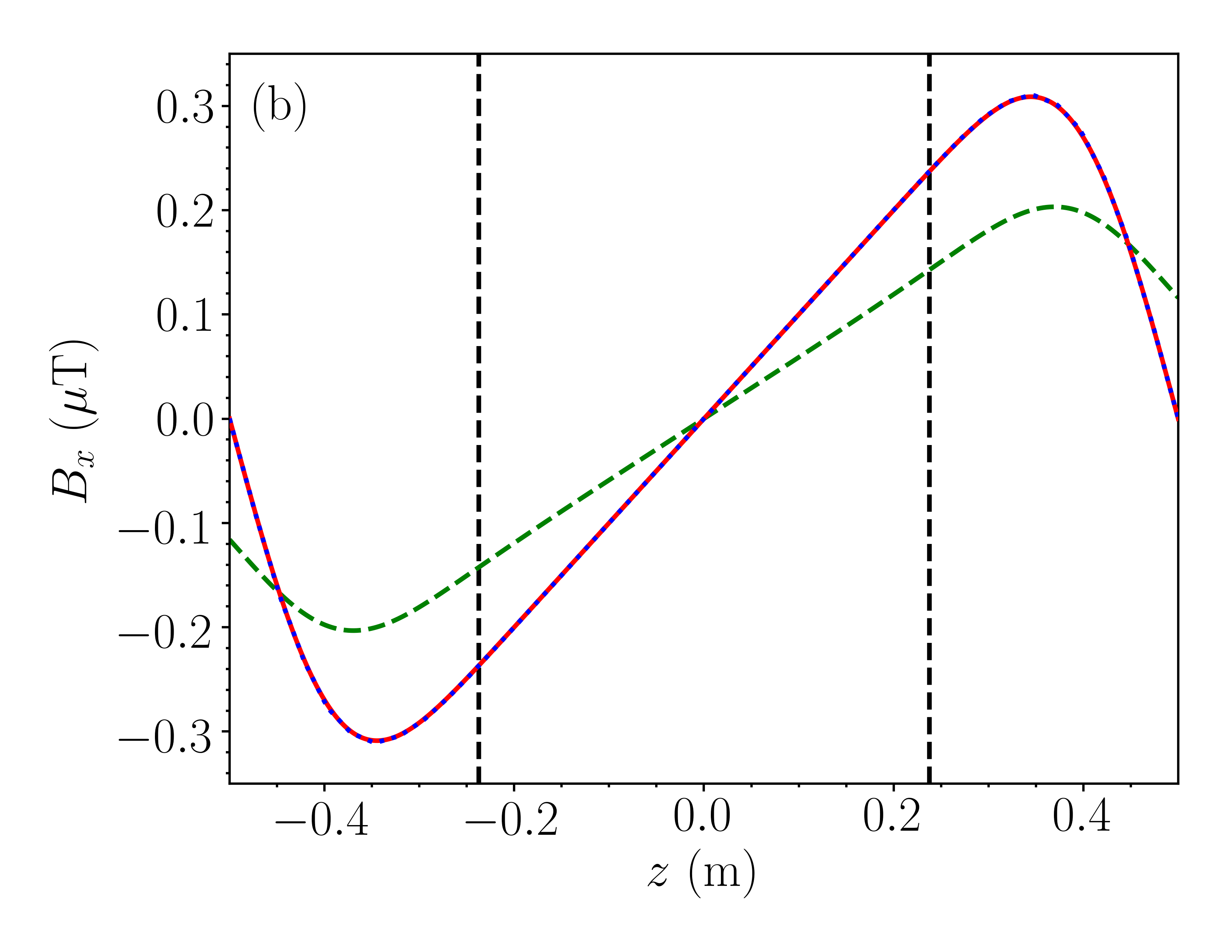}}
        \caption{Wire layouts (a) and performance (b) of an optimized hybrid active--passive linear transverse gradient field-generating system in which current flows on a cylinder of length $L_c=0.95$ m and radius $\rho_c=0.245$ m. The wire layouts are optimized to generate a linear transverse field gradient, $\mathrm{d}B_x/\mathrm{d}z=1$ $\mu$T/m, along the $z$-axis of the cylinder. The current-carrying cylinder is placed symmetrically inside a perfect closed magnetic shield of length $L_s=1$ m and radius $\rho_s=0.25$ m and the magnetic field is optimized between $\rho=[0,\rho_c/2]$ and $z=\pm{L_c}/4$; dashed black lines in (b). The least squares optimization was performed with parameters $N=200$, $M=1$, and $\beta=5.95\times10^{-12}$. (a) Color map of the optimal current streamfunction on the cylinder [blue and red shaded regions correspond to the flow of current in opposite senses respectively and their intensity shows the streamfunction magnitude from low (white) to high (intense color)]. Solid and dashed black curves represent discrete wires with opposite senses of current flow, approximating the current continuum with $N_\varphi=24$ contour levels. (b) Transverse magnetic field, $B_x$, versus axial position, $z$, calculated from the current continuum in (a) in three ways: analytically using \eqref{eq.brf}-\eqref{eq.bzf} (solid red curve); numerically using COMSOL Multiphysics\textsuperscript{\textregistered} Version 5.3a and modelling the high-permeability cylinder as a perfect magnetic conductor (blue dotted curve); numerically \emph{without} the high-permeability cylinder and using the Biot–Savart law with $N_{\varphi}=100$ contour levels (dashed green).}
        \label{fig.golaynew}
\end{figure*}

In Fig.~\ref{fig.bxnew} and Fig.~\ref{fig.golaynew} respectively, we show active--passive systems for generating a constant transverse field, $B_x$, normal to the axis of the cylinder, and a linear transverse field gradient, $\mathbf{B}=(z~\mathbf{\hat{x}}+x~\mathbf{\hat{z}})$, along the axis of the cylinder. Each of these systems has a cylindrical surface of length $L_c=0.95$ m and radius $\rho_c=0.245$ m, which carries the coil current distribution. This current distribution is interior to and centred about the origin of a closed perfect magnetically conducting cylinder of length $L_s=1$ m and radius $\rho_s=0.25$ m. The field is optimized over the central cylindrical region spanning half the radius and length of the coil cylinder. Fig.~\ref{fig.bxnew}a and Fig.~\ref{fig.golaynew}a show the respective streamfunctions on the surface of the cylinder and their discretization into wire patterns. The magnetic fields shown in Fig.~\ref{fig.bxnew}b--c and Fig.~\ref{fig.golaynew}b are calculated in three ways: analytically, using our theoretical model in \eqref{eq.brf}-\eqref{eq.bzf}; numerically, using COMSOL Multiphysics\textsuperscript{\textregistered} with the shield treated as a perfect magnetic conductor; and numerically in free space, i.e. excluding the high-permeability material and calculating the magnetic fields by applying the elemental Biot--Savart law directly to the discrete coil geometries in Fig.~\ref{fig.bxnew}a and Fig.~\ref{fig.golaynew}a. It is clear from Figs.~\ref{fig.bxnew}--\ref{fig.golaynew} that our design methodology is capable of generating highly accurate user-specified target magnetic fields inside the optimized field region with good agreement between the theoretical model and numerical simulations.\\

We quantify this agreement by analyzing the deviation from the target fields in the optimization region, $\Delta{B_x}$ and $\Delta{\mathrm{d}B_x/\mathrm{d}z}$, for the constant transverse field and linear transverse field gradient systems, respectively. Specifically, along the $z$-axis of the optimized region the maximum absolute deviations from the target fields are $0.11$\% and $0.24$\%, respectively. Over the same region, the maximum absolute deviations between the numerically-simulated and analytically-calculated field profiles are $0.002$\% and $0.003$\%, respectively. We can also see the hybrid nature of our optimization by the improved performance of the active systems when inside, and coupled to, the passive shield. For example, the strength of the $B_x$ field is nearly doubled, and its uniformity is improved by a factor of $20$, when the high-permeability cylinder is added to the constant transverse field-generating system (see Fig.~\ref{fig.bxnew}b).\\

\begin{figure*}[!htb]
         {\includegraphics[scale=0.3185]{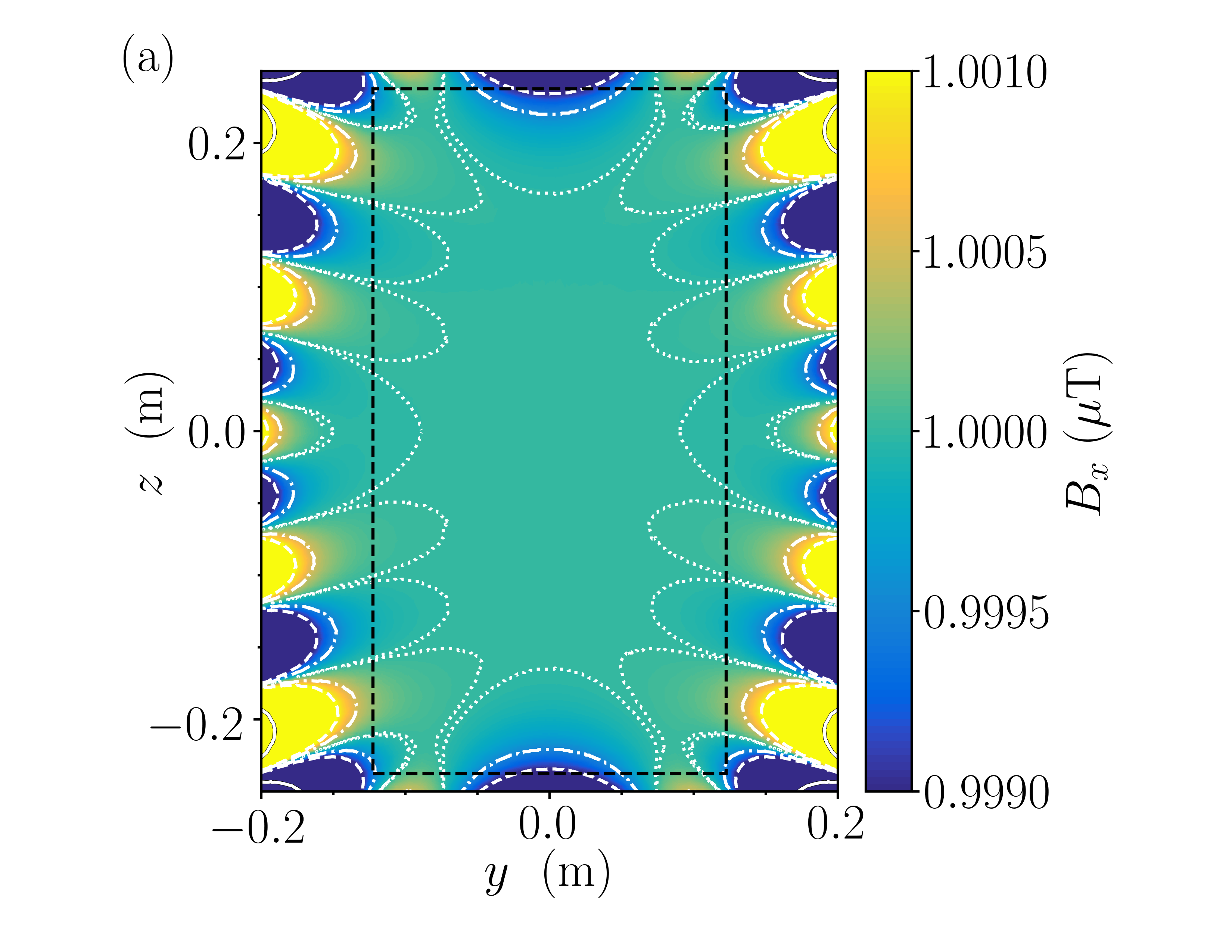}}
         {\includegraphics[scale=0.3185]{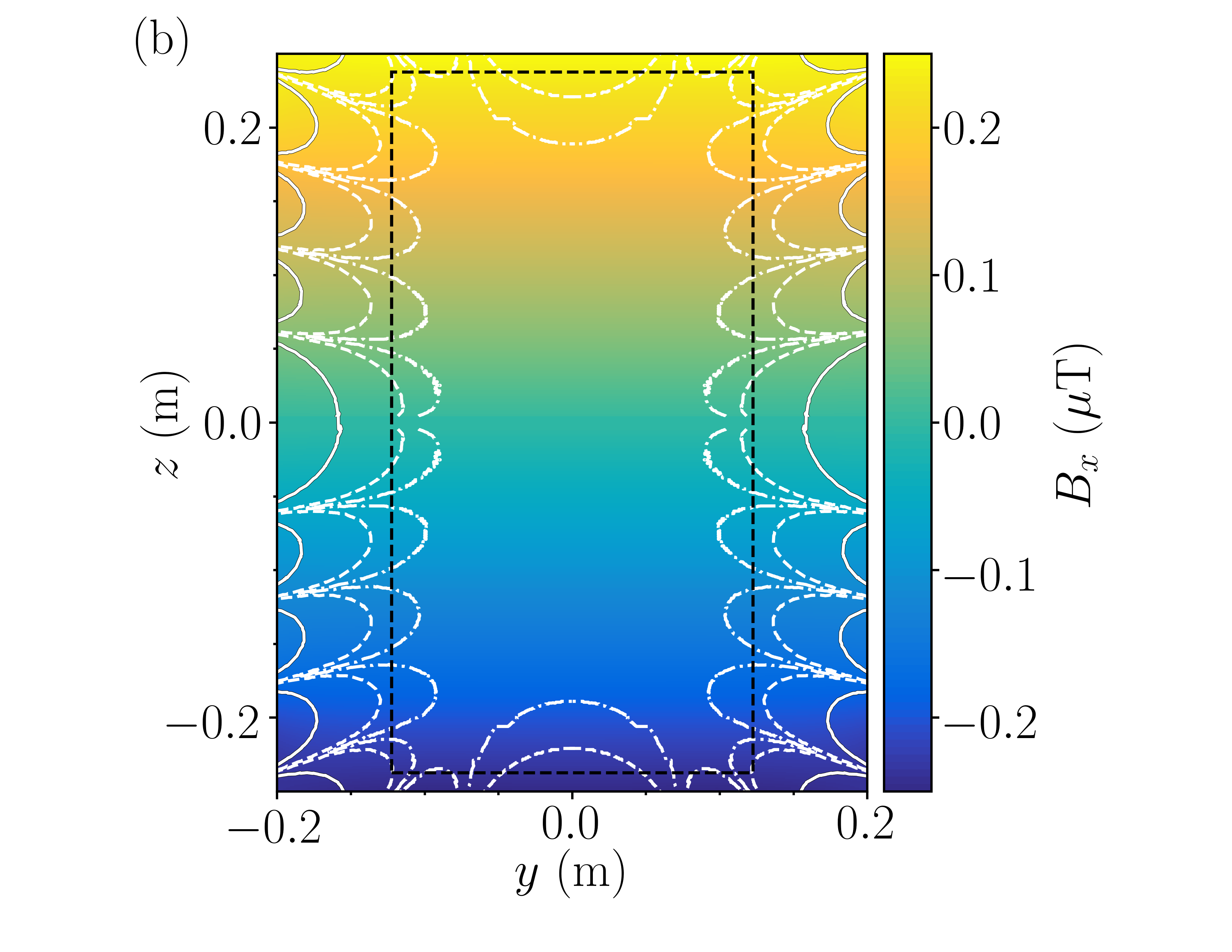}}
     \caption{Color maps showing the magnitude of the transverse magnetic field, $B_x$, in the $y$-$z$ plane inside a closed finite length perfect magnetic conductor generated by two active--passive systems: (a) the constant transverse field-generating system depicted in Fig.~\ref{fig.bxnew}; (b) the linear transverse gradient field-generating system depicted in Fig.~\ref{fig.golaynew}. The field profiles were calculated numerically using COMSOL Multiphysics\textsuperscript{\textregistered} Version 5.3a. Contours show where the field deviates from the target field by $1$\% (solid curves), $0.1$\% (dashed curves), $0.01$\% (dot-dashed curves), and $0.001$\% (dotted curves; in (a) only).}
     \label{fig.PMCcolormaps}
\end{figure*}

In Fig.~\ref{fig.PMCcolormaps} we show the numerically-calculated color maps of the $B_x$ field in the $y$-$z$ plane generated by the constant transverse field (Fig.~\ref{fig.bxnew}a) and linear transverse field gradient (Fig.~\ref{fig.golaynew}a) systems, using COMSOL Multiphysics\textsuperscript{\textregistered} with the shield treated as a perfect magnetic conductor. We summarize the performance of both systems in Table~\ref{tab.results}, calculating the cylindrical shield fractions -- the ratio of the radial and axial extent of the central region to that of the passive shield -- where the maximum deviations from the target fields are less than $0.01\%$, $0.05\%$, $0.1\%$, $0.5\%$, $1\%$, and $5\%$, respectively.\\ 

\begin{table}[!htb]
\renewcommand{\arraystretch}{1.75}
    \centering
    \begin{tabular}{c c c c c}
    \multicolumn{1}{c}{} & \multicolumn{4}{c}{\textbf{Cylindrical Shield Fraction}} \\
    \cline{2-5}
    \multicolumn{1}{c}{} & \multicolumn{2}{c}{\textbf{Perfect}} & \multicolumn{2}{c}{\textbf{Imperfect}} \\ 
    \hline
        \multicolumn{1}{c|}{\textbf{Max. Field Deviation (\%)}} & \multicolumn{1}{c}{$\mathbf{B_x}$} & \multicolumn{1}{c}{$\mathbf{\boldsymbol{\mathrm{d}}B_x/\boldsymbol{\mathrm{d}}z}$} & \multicolumn{1}{c}{$\mathbf{B_x}$} & \multicolumn{1}{c}{$\mathbf{\boldsymbol{\mathrm{d}}B_x/\boldsymbol{\mathrm{d}}z}$}  \\
        \hline
        \multicolumn{1}{c|}{$\mathbf{0.01}$} & $0.320$ & $0.237$ & $0.190$ & $0.168$ \\
        \multicolumn{1}{c|}{$\mathbf{0.05}$} & $0.398$ & $0.315$ & $0.350$ & $0.268$ \\
        \multicolumn{1}{c|}{$\mathbf{0.1}$} & $0.437$ & $0.348$ & $0.442$ & $0.320$ \\
        \multicolumn{1}{c|}{$\mathbf{0.5}$}  & $0.472$ & $0.448$ & $0.474$ & $0.440$ \\
        \multicolumn{1}{c|}{$\mathbf{1}$}  & $0.496$ & $0.470$ & $0.499$ & $0.470$ \\
        \multicolumn{1}{c|}{$\mathbf{5}$}  & $0.598$ & $0.553$ & $0.598$ & $0.551$ \\
        \\
    \end{tabular}
    
    \caption{Cylindrical shield fractions -- defined as the ratio of the radius and length of the central region to those of the passive shield -- where the maximum magnetic field deviations are within $0.01\%$, $0.05\%$, $0.1\%$, $0.5\%$, $1\%$, and $5\%$ of the target fields generated by the constant transverse, $B_x$, and linear transverse gradient, $\mathrm{d}B_x/\mathrm{d}z$, systems, depicted in Fig.~\ref{fig.bxnew}a and Fig.~\ref{fig.golaynew}a, respectively. The magnetic field deviations were calculated numerically using COMSOL Multiphysics\textsuperscript{\textregistered} Version 5.3a in two ways: inside a perfect magnetic conductor (perfect case); inside a magnetic shield with finite permeability $\mu_r=20000$, thickness $d=1$ mm, and a circular entry hole of normalized radius, $\rho_h=0.25\rho_s$ in both end caps (imperfect case).}
    \label{tab.results}
\end{table}
When physically constructing active--passive structures for real-world experiments, additional limitations must be taken into consideration in order to generate accurate magnetic fields using our design methodology. These limitations originate either from the theoretical model itself or from experimental practicalities. The limitations in the theoretical model are primarily associated with how accurately the high-permeability cylinder approximates a perfect magnetic conductor. This depends on the value of the finite permeability, the thickness of the shielding material, and the required experimental access holes in the shielding system. The errors introduced by these parameters depend on the lengths, radii, and positions of the conducting and high-permeability cylinders relative to the location of the optimization region. The experimental limitations on the field fidelity relate to the stability of the experimental equipment and errors in manufacturing an accurate representation of the current continuum. These errors include coarse discretization of the current continuum, inexact wire placement and construction, and imprecise positioning of the active structure inside the high-permeability shield. In practice, highly stable experimental equipment is available, particularly power supplies and current drivers \cite{Ac_n_2018}, and it is possible to manufacture structures that approximate the current continuum accurately~\cite{turner}. Consequently, the error introduced in the formulation of our model must be calculated to determine the accuracy of any future potential designs in a real-world experimental set-up. Our active--passive systems must, therefore, be analyzed for the case of a high-permeability cylinder that is not a closed perfect magnetic conductor. To do this, we use COMSOL Multiphysics\textsuperscript{\textregistered} working in the magnetostatic regime, to determine how the uniformity of the $B_x$ field generated by the constant transverse field-generating system in Fig.~\ref{fig.bxnew}a changes when the perfect closed magnetic conductor is replaced by one that is imperfect. To construct an imperfect shield, we define the magnetic permeability $\mu_r=20000$ to match the permeability of industrial standard mumetal regularly used as a passive shield. We then vary the wall thickness, $d$, of a closed magnetic shield and determine how small $d$ can be while maintaining high field uniformity. Finally, using this minimum thickness, we introduce a circular axial entry hole of radius $\rho_h$ at the center of each end cap and determine how large the hole can be to preserve field uniformity.\\

We use the root mean square field deviation, ${\Delta}B_x^\mathrm{RMS}$, of the attained field from the uniform target field, calculated along the $z$-axis of the optimized field region, to evaluate the performance of the active--passive system. The light blue crosses in Fig.~\ref{fig.physical}a show that ${\Delta}B_x^\mathrm{RMS}$ values calculated numerically from the current continuum in Fig.~\ref{fig.bxnew}a using COMSOL Multiphysics\textsuperscript{\textregistered}, decrease as $d$ increases in the range $(0.05-2.5)$ mm with interval size $0.05$ mm, converging to the \emph{thick} material limit where the thickness is assumed to be infinite. The horizontal dashed red line shows the analytical value of ${\Delta}B_x^\mathrm{RMS} = 0.0232 \%$ calculated using \eqref{eq.brf}-\eqref{eq.bzf}. The numerical ${\Delta}B_x^\mathrm{RMS}$ values decrease asymptotically below this analytical limit and approach the difference $\mathcal{O}(\mu_r^{-1})\approx0.005\%$~\cite{mu,mu1} that we predicted for our model in Section~\ref{sec.theory}. This intrinsic error, resulting from the small difference between \emph{thick} high-permeability materials and a perfect magnetic conductor, sets the hard limit on the accuracy of any magnetic field that can by designed using our theoretical model. In reality, however, this limit is so small that for a thick material with a high permeability, such as mumetal, the errors in manufacturing and construction will be much more significant. As technologies advance which reduce these system errors, such as screen-printed foldable PCBs~\cite{PCBCoils} and 3D-printing technologies~\cite{3DPrintingPaper}, it may become more relevant to develop a model which accounts \emph{ab initio} for magnetic shields of finite permeability and thickness.\\ 

We see from Fig.~\ref{fig.physical}a that the asymptotic limit is reached at approximately $d=1$ mm, where ${\Delta}B_x^\mathrm{RMS}=0.019\%$. At this point, regarding the accuracy of our model, there is little advantage to increasing $d$ further. Consequently, in Fig.~\ref{fig.physical}b we take $d=1$ mm and examine the effect of introducing circular axial entry holes in both end caps of the high-permeability cylinder. Although ${\Delta}B_x^\mathrm{RMS}$ increases as the hole radius, $\rho_h$, increases from no hole, $\rho_h=0$, to when there is no end cap, $\rho_h=\rho_s$, we see that small holes in both end caps allow experimental access without significantly reducing the fidelity of the desired field profile. In particular, for our system, the hole radius can be made as large as $\rho_h=0.25\rho_s$ while only increasing ${\Delta}B_x^\mathrm{RMS}$ by $0.0012\%$ when compared to the no hole case (horizontal dashed light blue line).\\

In Fig.~\ref{fig.finalcolormaps} we show the numerically-calculated color maps of the $B_x$ field in the $y$-$z$ plane generated by the constant transverse field (Fig.~\ref{fig.bxnew}a) and linear transverse field gradient (Fig.~\ref{fig.golaynew}a) systems. Both of these systems are simulated with the same imperfect high-permeability cylindrical magnetic shield that has had its properties determined in the above analysis: $\mu_r=20000$, $d=1$ mm, and $\rho_h=0.25\rho_s$. The performance of both systems in terms of the cylindrical shield fraction is summarized in Table~\ref{tab.results}.\\

Finally, we see from Table~\ref{tab.results} that this imperfect magnetic shield does not introduce significant magnetic field deviations above $0.1\%$ when compared to a perfect shield with the same geometry. In particular, the maximum difference between the perfect and imperfect cylindrical shield fractions for deviations above $0.1\%$ is only $0.028$. Large deviations in the field accuracy below $0.1\%$ can be graphically seen when comparing the color maps for the perfect (Fig.~\ref{fig.PMCcolormaps}) and imperfect (Fig.~\ref{fig.finalcolormaps}) cases. The contours showing field deviations of $0.01\%$ and $0.001\%$ are strongly perturbed, as expected from the analysis in Fig.~\ref{fig.physical}, demonstrating the \emph{hard} intrinsic limit on our model when generating target field profiles inside real-world magnetic shields. Similar analysis should be applied when designing other active--passive systems using our theoretical model in order to quantify its accuracy for a specific experimental setup. Further analysis could also be performed to determine the low-frequency limit in which a time-dependent current source could be included. However, if the magnitudes of any induced eddy currents are much less than the magnitude of the coil current, such effects will be negligible. Further designs generating more exotic field profiles can be found in the Supplementary Material. All designs can be found in our open access Python code which can be used to design systems to generate specific physical target field using our model (see addendum for more details).

\begin{figure}[!htb]
        {\begin{tikzpicture}[      
            every node/.style={anchor=south west,inner sep=0pt},
            x=1cm, y=1cm,
            ]   
            \node (fig1) at (0,0)
            {\includegraphics[scale=0.3185]{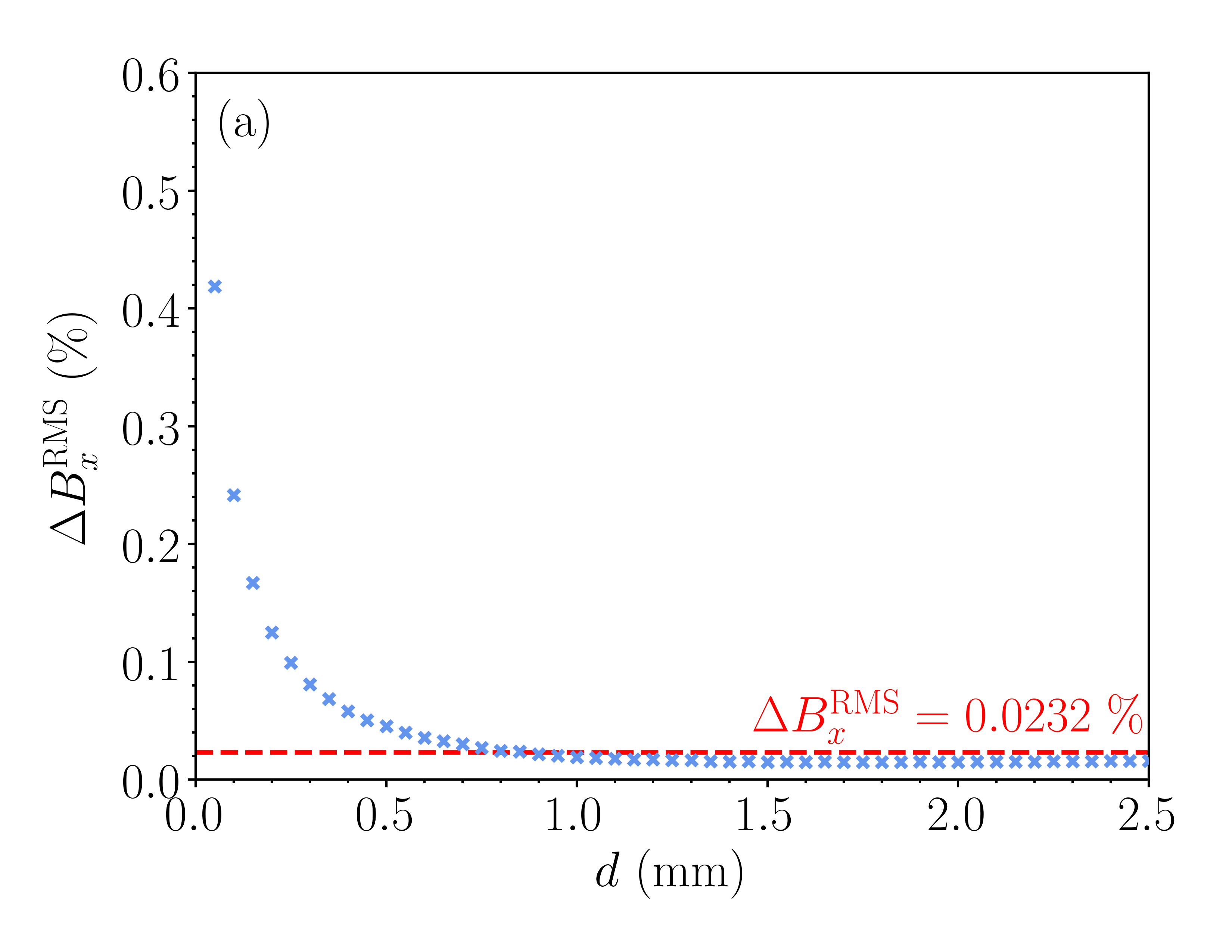}};
            \node (fig2) at (6.2,2)
            {\begin{tikzpicture}[scale=0.32]
            \draw (-1.25,0) arc (180:360:1.25 and -0.5);
            \draw (-1.25,0) arc (180:360:1.25 and 0.5);
            \draw (-1.25,0) -- (-1.25,-3.5);
            \draw (-1.25,-3.5) arc (180:360:1.25 and 0.5);
            \draw (-1.25,-3.5) arc (180:360:1.25 and -0.5);
            \draw (1.25,-3.5) -- (1.25,0);  
            \draw (0,0.5) ellipse (1.75 and 0.75);
            \draw (-1.75,0.5) -- (-1.75,-4);
            \draw (1.75,0.5) -- (1.75,-4);
            \draw (-1.75,-4) arc (180:360:1.75 and 0.75);
            \draw (-1.75,-4) arc (180:360:1.75 and -0.75);
            \draw [->] (0,-2) -- (-1.25,-2);
            \draw [<-] (-1.75,-2) -- (-2.5,-2);
            \node at (-1,-1.5) {$d$}; 
            \end{tikzpicture}};  
        \end{tikzpicture}}
        {\begin{tikzpicture}[      
            every node/.style={anchor=south west,inner sep=0pt},
            x=1cm, y=1cm,
            ]   
            \node (fig1) at (0,0)
            {\includegraphics[scale=0.3185]{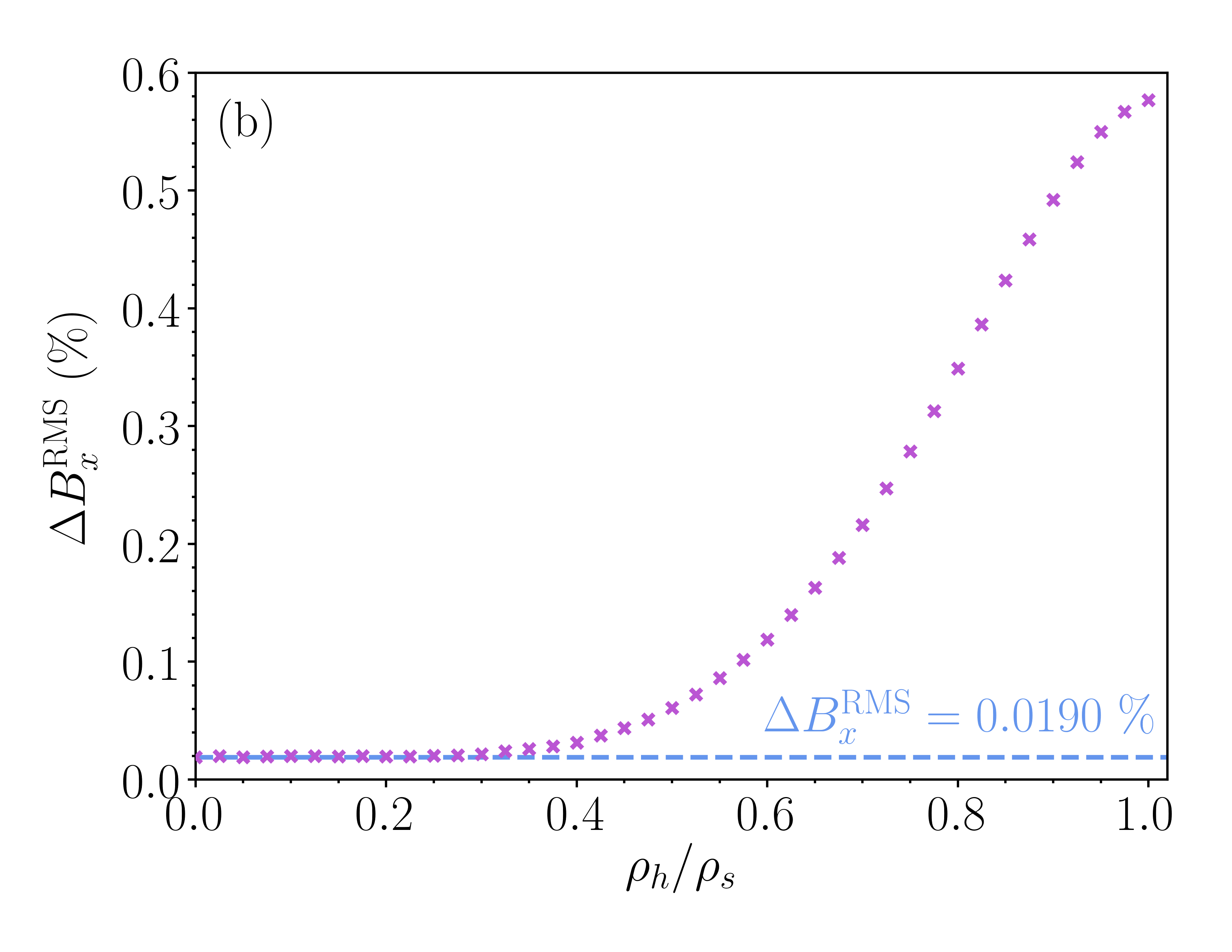}};
            \node (fig2) at (6.5,2)
            {\begin{tikzpicture}[scale=0.37]
            \draw (-1.25,0) arc (180:360:1.25 and 1.25);
            \draw (-1.25,0) arc (180:360:1.25 and -1.25);
            \draw (-1.75,0) arc (180:360:1.75 and 1.75);
            \draw (-1.75,0) arc (180:360:1.75 and -1.75);
            \draw [->] (0,0) -- (-1.25,0);
            \node at (-1,0.25) {$\rho_h$}; 
            \end{tikzpicture}};  
        \end{tikzpicture}}
    \caption{Plots showing how the performance of the optimized hybrid active--passive constant transverse field-generating system depicted in Fig.~\ref{fig.bxnew} depends (a) on the thickness, $d$, of the high-permeability cylinder (inset), and (b) on the radius, $\rho_h$, of the circular axial entry hole in the two end caps (inset), which allow access to the interior of the system. (a) Root mean square (RMS) deviation, ${\Delta}B_x^\mathrm{RMS}$, evaluated along the axis of the optimized field region, as a function of the high-permeability shield thickness, $d$, taking $\mu_r=20000$ and $\rho_h=0$. Light blue crosses show ${\Delta}B_x^\mathrm{RMS}$ values calculated numerically from the current continuum in Fig.~\ref{fig.bxnew}a using COMSOL Multiphysics\textsuperscript{\textregistered} Version 5.3a. Horizontal dashed red line shows the analytical value of ${\Delta}B_x^\mathrm{RMS} = 0.0232 \%$ calculated using \eqref{eq.brf}-\eqref{eq.bzf}. (b) ${\Delta}B_x^\mathrm{RMS}$ evaluated along the axis of the optimized field region, as a function of the normalized axial hole radius, $\rho_h/\rho_s$, taking $\mu_r=20000$ and $d=1$. Purple crosses show ${\Delta}B_x^\mathrm{RMS}$ values calculated numerically from the current continuum in Fig.~\ref{fig.bxnew}a using COMSOL Multiphysics\textsuperscript{\textregistered} Version 5.3a. For comparison, the horizontal dashed light blue line shows the numerical RMS deviation ${\Delta}B_x^\mathrm{RMS}= 0.0190 \%$ calculated when $d=1$ mm and $\rho_h=0$.}
    \label{fig.physical}
\end{figure}
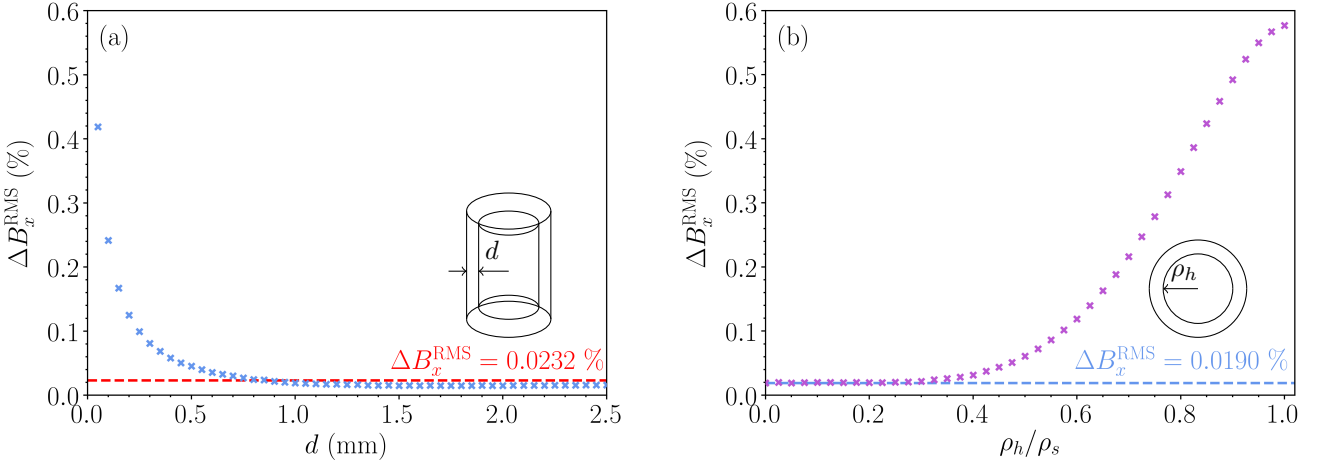
\begin{figure}[!htb]
         {\includegraphics[scale=0.3185]{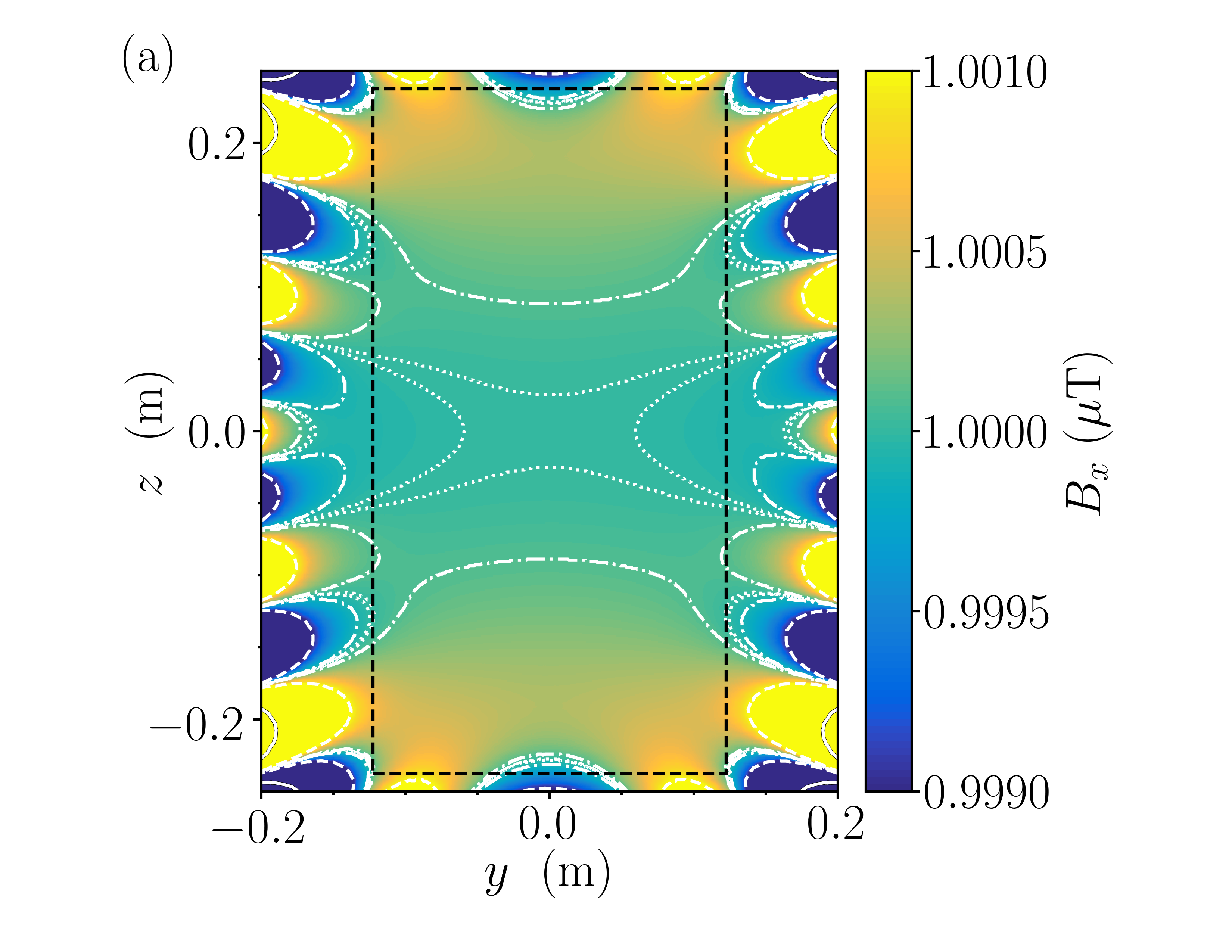}}
         {\includegraphics[scale=0.3185]{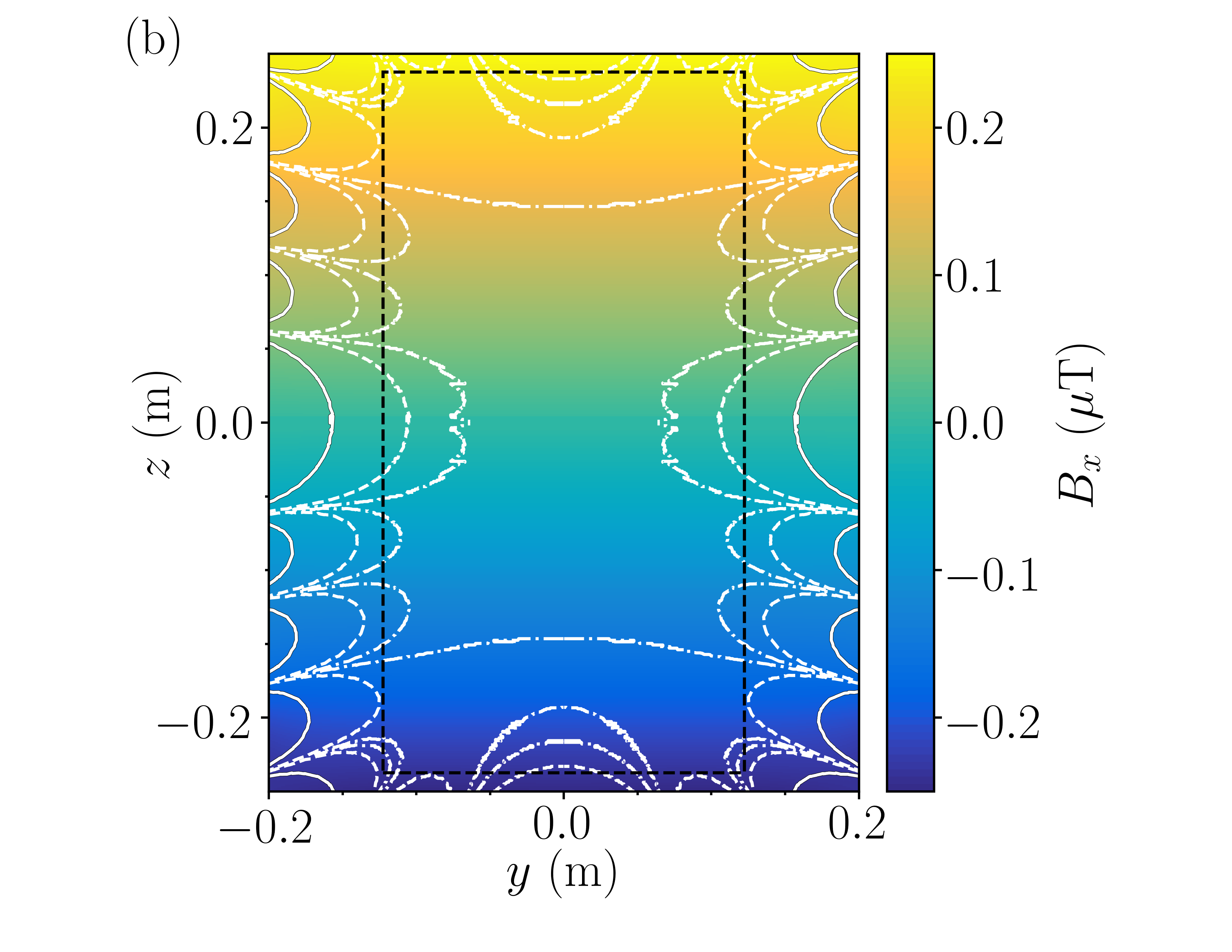}}
     \caption{Color maps showing the magnitude of the transverse magnetic field, $B_x$, in the $y$-$z$ plane inside a magnetic shield with permeability $\mu_r=20000$, thickness $d=1$ mm, and a circular axial entry hole of normalized radius, $\rho_h=0.25\rho_s$, around the centre of each end cap generated by two active--passive systems: (a) the constant transverse field-generating system depicted in Fig.~\ref{fig.bxnew}; (b) the linear transverse gradient field-generating system depicted in Fig.~\ref{fig.golaynew}. The field profiles were calculated numerically using COMSOL Multiphysics\textsuperscript{\textregistered} Version 5.3a. Contours show where the field deviates from the target field by $1$\% (solid curves), $0.1$\% (dashed curves), $0.01$\% (dot-dashed curves), and $0.001$\% (dotted curves; in (a) only).}
     \label{fig.finalcolormaps}
\end{figure}
\section{Conclusion}
In this paper, we have developed an analytical model to calculate the total magnetic field generated by an arbitrary current flow on a cylinder that is coaxially nested within a finite closed high-permeability cylinder. We modified the Green's function for the magnetic vector potential, matched the radial and planar boundary conditions of the magnetic field through the introduction of a pseudo-current density, and incorporated a harmonic minimization procedure to design optimal user-specified magnetic fields by using a modified cylindrical Fourier basis. We then verified this optimization procedure by designing coils to generate a constant transverse field, $B_x$, across the cylinder, and a linear transverse field gradient, $\mathrm{d}B_x/\mathrm{d}z$, along the length of the cylinder. Our analytical calculations of these field profiles agreed well with numerical simulations. The optimization procedure generated highly accurate $B_x$ and $\mathrm{d}B_x/\mathrm{d}z$ field profiles, inside a high-permeability cylinder, with peak-to-peak deviations from the target profiles below $0.11\%$ and $0.24\%$, respectively. The analytically-predicted deviations agreed with the numerical simulations to within $0.002\%$ and $0.003\%$ for the constant and linear gradient systems, respectively.\\

We further investigated the validity of our model by analyzing the behavior of the constant transverse field-generating system inside a magnetic shield of permeability $\mu_r=20000$, finite thickness, and with circular axial entry holes in the end caps. We found a range of parameters where the analytical predictions for a perfect cylindrical magnetic conductor remain close to numerical simulations for a cylindrical shield with finite permeability and thickness that includes entry holes; showing that the designs generated by our model are applicable to real-world magnetic shields. Notably, when the active field-generating systems were enclosed by a passive magnetic shield with realistic experimental parameters ($\mu_r=20000$, thickness $1$ mm, and entry holes of radius equal to $25$\% of the shield's radius), the deviation from the desired constant and linear gradient field profiles were less than $0.1\%$ and $0.5\%$, respectively, over more than $40\%$ of the central radial and axial extent of this simulated real-world magnetic shield.\\ 

Our flexible optimization procedure enables the design of new active–passive magnetic field shaping systems to generate accurately any static magnetic field profile in the interior of a finite closed magnetic shield, subject to satisfying Maxwell’s equations in free space and not saturating the shielding material. This facilitates the development and miniaturization of systems and technologies which require such control, including quantum sensors, fundamental physics experiments, and medical technologies. Further investigation could consider an analytical treatment of finite magnetic shield thickness and permeability and interactions with an open magnetic shield topology.

\section*{Addendum}
\textbf{Acknowledgements} \quad We acknowledge support from the UK Quantum Technology Hub for Sensing and Timing, funded by the Engineering and Physical Sciences Research Council (EP/M013294/1). \\
\textbf{Competing Interests} \quad The authors M.~Packer, P.~J.~Hobson, T.~M.~Fromhold, M.~J.~Brookes, and R.~Bowtell declare that they have a patent pending to the UK Government Intellectual Property Application office (Application Number 1913549.0) regarding the magnetic field optimization techniques described in this work. The authors have no other competing financial interests. \\
\textbf{Correspondence} \quad
Correspondence and requests should be addressed to Mark.Fromhold@nottingham.ac.uk. \\
\textbf{Open Access} \quad
The Python code used to design arbitrary cylindrical coils in a magnetically shielded cylinder can be found in the following GitHub repository: https://github.com/peterjhobson/fields{\textunderscore}in{\textunderscore}shields. Verification using COMSOL Multiphysics\textsuperscript{\textregistered} requires a valid license.


\begin{thebibliography}{}

\end{thebibliography}


\begin{thebibliography}{10}

\bibitem{Wueaax0800}
X. Wu, Z. Pagel, B.~S. Malek, T.~H. Nguyen, F. Zi, D.~S.
  Scheirer, and H. M{\"u}ller.
\newblock Gravity surveys using a mobile atom interferometer.
\newblock {\em Sci. Adv.}, 5(9), 2019.

\bibitem{doi:10.1038/s41598-018-30608-1}
V. M{e}noret, P. Vermeulen, N.~L. Moigne, S. Bonvalot,
  P. Bouyer, A. Landragin, and B. Desruelle.
\newblock Gravity measurements below {$10^{-9}$} $g$ with a transportable
  absolute quantum gravimeter.
\newblock {\em Sci. Rep.}, 8(12300), 2018.

\bibitem{SnaddenGradiometer}
M.~J. Snadden, J.~M. McGuirk, P.~Bouyer, K.~G. Haritos, and M.~A. Kasevich.
\newblock Measurement of the {Earth's} gravity gradient with an atom
  interferometer-based gravity gradiometer.
\newblock {\em Phys. Rev. Lett.}, 81(5):971, 1998.

\bibitem{AIQuantumSensors}
K. Bongs, M. Holynski, J. Vovrosh, P. Bouyer, G. Condon,
  E. Rasel, C. Schubert, W.~P. Schleich, and A. Roura.
\newblock Taking atom interferometric quantum sensors from the laboratory to
  real-world applications.
\newblock {\em Nat. Rev. Phys.}, Oct. 2019.

\bibitem{PhysRevLett.113.013001}
B. Patton, E. Zhivun, D.~C. Hovde, and D. Budker.
\newblock All-optical vector atomic magnetometer.
\newblock {\em Phys. Rev. Lett.}, 113:013001, Jul. 2014.

\bibitem{quspin}
Y.~J. Kim and I. Savukov.
\newblock Ultra-sensitive magnetic microscopy with an optically pumped
  magnetometer.
\newblock {\em Sci. Rep.}, 6(24773), 2016.

\bibitem{10.1155/2015/491746}
H. Zhang, S. Zou, X. Chen, and W. Quan.
\newblock Parameter modeling analysis and experimental verification on magnetic
  shielding cylinder of all-optical atomic spin magnetometer.
\newblock {\em J. Sens.}, 2015:1--7, 05 2015.

\bibitem{ROMALIS2011258}
M.~V. Romalis and H.~B. Dang.
\newblock Atomic magnetometers for materials characterization.
\newblock {\em 	Mater. Today}, 14(6):258--262, 2011.

\bibitem{10.1364/BOE.3.000981}
T. Sander, J. Preusser, R. Mhaskar, J. Kitching, L. Trahms, and S. Knappe.
\newblock Magnetoencephalography with a chip-scale atomic magnetometer.
\newblock {\em Biomed. Opt. Express}, 3:981--90, 05 2012.

\bibitem{nature}
E. Boto, N. Holmes, J. Leggett, G. Roberts, V. Shah, S.~S. Meyer, L.~D. Muñoz,
K.~J. Mullinger, T.~M. Tierney, S. Bestmann, G.~R. Barnes, R. Bowtell, and M.~J. Brookes.
\newblock Moving magnetoencephalography towards real-world applications with a
  wearable system.
\newblock {\em Nature}, 555(657-661), 2018.

\bibitem{doi:10.1063/1.4919366}
I.~Altarev, M.~Bales, D.~H. Beck, T.~Chupp, K.~Fierlinger, P.~Fierlinger,
  F.~Kuchler, T.~Lins, M.~G. Marino, B.~Niessen, G.~Petzoldt, U.~Schläpfer,
  A.~Schnabel, J.~T. Singh, R.~Stoepler, S.~Stuiber, M.~Sturm, B.~Taubenheim,
  and J.~Voigt.
\newblock A large-scale magnetic shield with {$10^6$} damping at millihertz
  frequencies.
\newblock {\em J. Appl. Phys.}, 117(18):183903, 2015.

\bibitem{10.1007/s10751-014-1109-5}
Y.~Sakamoto, C.~Bidinosti, Y.~Ichikawa, Takefumi Satoh, Y.~Ohtomo, S.~Kojima,
  C.~Funayama, T.~Suzuki, M.~Tsuchiya, T.~Furukawa, A.~Yoshimi, T.~Ino,
  H.~Ueno, Y.~Matsuo, T.~Fukuyama, and K.~Asahi.
\newblock Development of high-homogeneity magnetic field coil for {$^{129}$Xi}
  EDM experiment.
\newblock {\em Hyperfine Interact.}, 230, 04 2015.

\bibitem{MORIC2014287}
I. Moric, P. Laurent, P. Chatard, C.-M. de Graeve,
  S. Thomin, V. Christophe, and O. Grosjean.
\newblock Magnetic shielding of the cold atom space clock PHARAO.
\newblock {\em Acta Astronaut.}, 102:287--294, 2014.

\bibitem{LiangClock}
L. Liu, D.-S. Lü, W.-B. Chen, T. Li, Q.-Z. Qu, B. Wang, L. Li,
  W. Ren, Z.-R. Dong, J.-B. Zhao, W.-B. Xia, X. Zhao, J.-W. Ji,
  M.-F. Ye, Y.-G. Sun, Y.-Y. Yao, D. Song, Z.-G. Liang,
  S.-J. Hu, and Y.-Z. Wang.
\newblock In-orbit operation of an atomic clock based on laser-cooled
  {$^{87}$Rb} atoms.
\newblock {\em Nature Comms.}, 9, 12 2018.

\bibitem{RevModPhys.62.541}
N.~F. Ramsey.
\newblock Experiments with separated oscillatory fields and hydrogen masers.
\newblock {\em Rev. Mod. Phys.}, 62:541--552, Jul. 1990.

\bibitem{HindsEDM}
J.~J. Hudson, D.~M. Kara, I.~J. Smallman, B.~E. Sauer, M.~R. Tarbutt, and E.~A.
  Hinds.
\newblock Improved measurement of the shape of the electron.
\newblock {\em Nature}, 473:493--496, 2011.

\bibitem{ACMEEDM}
{The ACME Collaboration}.
\newblock Improved limit on the electric dipole moment of the electron.
\newblock {\em Nature}, 562:355--360, 2018.

\bibitem{doi:10.1063/1.4922671}
I.~Altarev, P.~Fierlinger, T.~Lins, M.~G. Marino, B.~Nießen, G.~Petzoldt,
  M.~Reisner, S.~Stuiber, M.~Sturm, J.~Taggart~Singh, B.~Taubenheim, H.~K.
  Rohrer, and U.~Schläpfer.
\newblock Minimizing magnetic fields for precision experiments.
\newblock {\em J. Appl. Phys.}, 117(23):233903, 2015.

\bibitem{PhysRevLett.105.151604}
J.~M. Brown, S.~J. Smullin, T.~W. Kornack, and M.~V. Romalis.
\newblock New limit on Lorentz- and CPT-violating neutron spin interactions.
\newblock {\em Phys. Rev. Lett.}, 105:151604, Oct. 2010.

\bibitem{PhysRevA.100.010501}
M.~E. Limes, N.~Dural, M.~V. Romalis, E.~L. Foley, T.~W. Kornack, A.~Nelson,
  L.~R. Grisham, and J.~Vaara.
\newblock Dipolar and scalar
  $^{3}\mathrm{He}\text{\ensuremath{-}}^{129}\mathrm{Xe}$ frequency shifts in
  stemless cells.
\newblock {\em Phys. Rev. A}, 100:010501, Jul. 2019.

\bibitem{PhysRevLett.112.110801}
F.~Allmendinger, W.~Heil, S.~Karpuk, W.~Kilian, A.~Scharth, U.~Schmidt,
  A.~Schnabel, Yu. Sobolev, and K.~Tullney.
\newblock New limit on lorentz-invariance- and $cpt$-violating neutron spin
  interactions using a free-spin-precession
  $^{3}\mathrm{He}$-$^{129}\mathrm{Xe}$ comagnetometer.
\newblock {\em Phys. Rev. Lett.}, 112:110801, Mar. 2014.

\bibitem{10.1063/1.1656455}
A.~Mager.
\newblock Magnetic shielding efficiencies of cylindrical shells with axis
  parallel to the field.
\newblock {\em J. Appl. Phys.}, 39, 02 1968.

\bibitem{GRABCHIKOV201649}
S.~S. Grabchikov, A.~V. Trukhanov, S.~V. Trukhanov, I.~S. Kazakevich, A.~A. Solobay,
  V.~T. Erofeenko, N.~A. Vasilenkov, O.~S. Volkova, and A. Shakin.
\newblock Effectiveness of the magnetostatic shielding by the cylindrical
  shells.
\newblock {\em J. Magn. Magn. Mater.}, 398:49--53,
  2016.

\bibitem{6217348}
J. Prat-Camps, C. Navau, D. Chen, and A. Sanchez.
\newblock Exact analytical demagnetizing factors for long hollow cylinders in
  transverse field.
\newblock {\em IEEE Magn. Lett.}, 3:0500104--0500104, 2012.

\bibitem{RandH}
F. Roméo and D.~I. Hoult.
\newblock Magnet field profiling: analysis and correcting coil design.
\newblock {\em Magn. Reson. Med.}, 1(1):44-65, Mar. 1984.

\bibitem{Pissanetzky_1992}
\newblock S. Pissanetzky.
\newblock Minimum energy {MRI} gradient coils of general geometry.
\newblock {\em Meas. Sci. Technol.}, 3(7):667--673, Jul. 1992.

\bibitem{doi:10.1002/cmr.b.20091}
\newblock M. Poole and R. Bowtell.
\newblock Novel gradient coils designed using a boundary element method.
\newblock {\em Concept. Magn. Reson. B}, 31B(3):162--175, 2007.

\bibitem{doi:10.1002/cmr.b.20040}
\newblock R.A. Lemdiasov and R. Ludwig.
\newblock A stream function method for gradient coil design.
\newblock {\em Concept. Magn. Reson. B}, 26B(1):67--80, 2005.

\bibitem{Forbes_2002}
\newblock L.~K. Forbes and S. Crozier.
\newblock A novel target-field method for finite-length magnetic resonance shim coils: {II}. Tesseral shims.
\newblock {\em J. Phys. D: Appl. Phys.}, 35, 9, 839--849 2002.

\bibitem{Forbes_2001}
\newblock L.~K. Forbes and S. Crozier.
\newblock A novel target-field method for finite-length magnetic resonance shim coils: {I}. Zonal shims.
\newblock {\em J. Phys. D: Appl. Phys.}, 34, 24, 3447--3455 2001.

\bibitem{niall0}
\newblock N. Holmes, T.~M. Tierney, J. Leggett, E. Boto, S. Mellor, 
G. Roberts, R.~M. Hill, V. Shah, G.~R. Barnes, M.~J. Brookes, and R. Bowtell.
\newblock Balanced, bi-planar magnetic field and field gradient coils for field compensation in wearable magnetoencephalography. 
\newblock {\em Sci. Rep.} 9, 14196 (2019).

\bibitem{doi:10.1002/9780470268483.app2}
S. Celozzi, R. Araneo, and G. Lovat.
\newblock {\em Appendix B: Magnetic Shielding}.
\newblock John Wiley \& Sons, Ltd, 2008.

\bibitem{genetic}
{S. Celozzi and F. Garzia}.
\newblock Active shielding for power-frequency magnetic field reduction using
  genetic algorithms optimisation.
\newblock {\em IEE Proceedings A - Sci., Mes., and Technol.},
  151(1):2--7, 2004.

\bibitem{genetic1}
{K.~F. Man, K.~S. Tang and S. Kwong}.
\newblock Genetic algorithms: concepts and applications [in engineering
  design].
\newblock {\em IEEE Trans. Ind.}, 43(5):519--534,
  1996.

\bibitem{CACIAGLI2018423}
A. Caciagli, R.~J. Baars, A.~P. Philipse, and B.~W.~M. Kuipers.
\newblock Exact expression for the magnetic field of a finite cylinder with
  arbitrary uniform magnetization.
\newblock {\em J. Magn. Magn. Mater.}, 456:423 -- 432,
  2018.

\bibitem{Solenoid1}
C.-Y. Liu, T.~Andalib, D.~C.~M. Ostapchuk, and C.~P. Bidinosti.
\newblock Analytic models of magnetically enclosed spherical and solenoidal
  coils.
\newblock {\em Nucl. Instrum. Methods Phys. Res.}, 949:162837,
  2020.

\bibitem{solenoid2}
R. Lambert and C. Uphoff.
\newblock Magnetically shielded solenoid with field of high homogeneity.
\newblock {\em Rev. Sci. Instrum.}, 46, 03 1975.

\bibitem{doi:10.1063/1.1719514}
R.~J. Hanson and F.~M. Pipkin.
\newblock Magnetically shielded solenoid with field of high homogeneity.
\newblock {\em Rev. Sci. Instrum.}, 36(2):179--188, 1965.

\bibitem{jackson}
J.~D. Jackson.
\newblock Classical electrodynamics (3rd ed.).
\newblock {\em Wiley}, New York, 1998.

\bibitem{cubefem}
{Q. Cao, D. Pan, J. Li, Y. Jin, Z. Sun, S. Lin, G.
  Yang, and L. Li}.
\newblock Optimization of a coil system for generating uniform magnetic fields
  inside a cubic magnetic shield.
\newblock {\em Energies}, 11(3):608, 2018.

\bibitem{niall}
{N. Holmes, J. Leggett, E. Boto, G. Roberts, R.~M. Hill, T.~M. Tierney, V. Shah, G.~R. Barnes, M.~J. Brookes, and R. Bowtell.}
\newblock A bi-planar coil system for nulling background magnetic fields in
  scalp mounted magnetoencephalography.
\newblock {\em NeuroImage}, 181(1):760--774, 2018.

\bibitem{8500866}
T. Liu, J. Voigt, Z. Sun, A. Schnabel, K. Grueneberg, I. Fan,
  and L. Li.
\newblock Two-step mirror model for the optimization of the magnetic field
  generated by coils inside magnetic shielding.
\newblock In {\em 2018 Conference on Precision Electromagnetic Measurements
  (CPEM 2018)}, 1--2, 2018.

\bibitem{LIU2020166846}
T. Liu, A. Schnabel, Z. Sun, J. Voigt, and L. Li.
\newblock Approximate expressions for the magnetic field created by circular
  coils inside a closed cylindrical shield of finite thickness and
  permeability.
\newblock {\em J. Magn. Magn. Mater.}, 507:166846, 2020.

\bibitem{greensey}
{L.-W. Li, M.-S. Leong, T.-S. Yeo, and P.-S. Kooi}.
\newblock Electromagnetic dyadic Green's functions in spectral domain for
  multilayered cylinders.
\newblock {\em J. Electromagnet. Wave}, 14:961--985,
  2000.

\bibitem{turner}
R. Turner.
\newblock A target field approach to optimal coil design.
\newblock {\em J. Phys. D: Appl. Phys.}, 19(8), 1986.

\bibitem{doi:10.1002/mrm.1910260202}
J.~W. Carlson, K.~A. Derby, K.~C. Hawryszko, and M.~Weideman.
\newblock Design and evaluation of shielded gradient coils.
\newblock {\em Magn. Reson. Med.}, 26(2):191--206, 1992.

\bibitem{hoult}
D.~I. Hoult and R.~Deslauriers.
\newblock Accurate shim-coil design and magnet-field profiling by a
  power-minimization-matrix method.
\newblock {\em J. Magn. Reson.}, 108:1(9--20), 1994.

\bibitem{mu}
{P. Hammond}.
\newblock Electric and magnetic images.
\newblock {\em Proc. IEE Part C Monogr.}, 107:306, 1960.

\bibitem{mu1}
{P. Hammond}.
\newblock Effect of finite thickness of magnetic substrate on planar inductors.
\newblock {\em IEEE Trans. Magn.}, 26:270--275, 1990.

\bibitem{ap}
R. Turner and R.~M. Bowley.
\newblock Passive screening of switched magnetic field gradients.
\newblock {\em J. Phys. E Sci. Instrum.}, 19(876), 1986.

\bibitem{Ac_n_2018}
A. Ac{\'{\i}}n, I. Bloch, H. Buhrman, T. Calarco,
  C. Eichler, J. Eisert, D. Esteve, N. Gisin, S.~J.
  Glaser, F. Jelezko, S. Kuhr, M. Lewenstein, M.~F. Riedel, P.~O.
  Schmidt, R. Thew, A. Wallraff, I. Walmsley, and F.~K. Wilhelm.
\newblock The quantum technologies roadmap: a {European} community view.
\newblock {\em New J. Phys.}, 20(8):080201, Aug. 2018.

\bibitem{PCBCoils}
J.~R. Corea, A.~M. Flynn, B. Lechene, G. Scott, G.~D. Reed,
  P.~J. Shin, M. Lustig, and A.~C. Ariasb.
\newblock Screen-printed flexible {MRI} receive coils.
\newblock {\em Nat. Commun.}, 7, Mar. 2016.

\bibitem{3DPrintingPaper}
T.~D. Ngo, A. Kashani, G. Imbalzano, K.~T.~Q. Nguyen, and D.
  Hui.
\newblock Additive manufacturing ({3D} printing): A review of materials,
  methods, applications and challenges.
\newblock {\em Compos. B. Eng.}, 143:172--196, 2018.

\end{thebibliography}
\bibliographystyle{apsrev4-2} 

\bibliographystyle{unsrt}

\newpage

\section*{Appendices}
\subsection*{A Mathematical Definitions} \label{app:maths}
\begin{align}\label{eq.fn}
    F_{n}\left(\rho,z\right)=\sum_{p=-\infty}^{\infty}\int_{-\infty}^\infty \mathrm{d}k\ I'_0\left(|k|\rho\right)\Bigg(K'_{0}(|k|\rho_c) -\frac{I'_{0}(|k|\rho_c)K_{0}(|k|\rho_s)}{I_{0}(|k|\rho_s)}\Bigg)C^1_{np}(k,z),
\end{align}
\begin{align}\label{eq.gn}
    G_{nm}\left(\rho,z\right)=\sum_{p=-\infty}^{\infty}\int_{-\infty}^\infty \mathrm{d}k\ I'_m\left(|k|\rho\right)\Bigg(K'_{m}(|k|\rho_c)-\frac{I'_{m}(|k|\rho_c)K_{m}(|k|\rho_s)}{I_{m}(|k|\rho_s)}\Bigg)C^2_{np}(k,z),
\end{align}
\begin{equation}
    G^w_{nm}\left(\rho,\phi,z\right)=\cos\left(m\phi\right)G_{nm}\left(\rho,z\right), \quad G^q_{nm}\left(\rho,\phi,z\right)=\sin\left(m\phi\right)G_{nm}\left(\rho,z\right),
\end{equation}
\begin{align}\label{eq.hn}
    H_{nm}\left(\rho,z\right)=\sum_{p=-\infty}^{\infty}\int_{-\infty}^\infty \mathrm{d}k\ mI_m\left(|k|\rho\right)\Bigg(K'_{m}(|k|\rho_c)-\frac{I'_{m}(|k|a)K_{m}(|k|\rho_s)}{I_{m}(|k|\rho_s)}\Bigg)C^3_{np}(k,z),
\end{align}
\begin{equation}
    H^w_{nm}\left(\rho,\phi,z\right)=\sin\left(m\phi\right)H_{nm}\left(\rho,z\right), \quad H^q_{nm}\left(\rho,\phi,z\right)=-\cos\left(m\phi\right)H_{nm}\left(\rho,z\right),
\end{equation}
\begin{align}\label{eq.dn}
    D_{n}\left(\rho,z\right)=\sum_{p=-\infty}^{\infty}\int_{-\infty}^\infty \mathrm{d}k\ I_0\left(|k|\rho\right)\Bigg(K'_{0}(|k|\rho_c)-\frac{I'_{0}(|k|\rho_c)K_{0}(|k|\rho_s)}{I_{0}(|k|\rho_s)}\Bigg)C^4_{np}(k,z),
\end{align}
\begin{align}\label{eq.sn}
    S_{nm}\left(\rho,z\right)=\sum_{p=-\infty}^{\infty}\int_{-\infty}^\infty \mathrm{d}k\ I_m\left(|k|\rho\right)\Bigg(K'_{m}(|k|\rho_c)-\frac{I'_{m}(|k|\rho_c)K_{m}(|k|\rho_s)}{I_{m}(|k|\rho_s)}\Bigg)C^5_{np}(k,z),
\end{align}
\begin{equation}
    S^w_{nm}\left(\rho,\phi,z\right)=\cos\left(m\phi\right)S_{nm}\left(\rho,z\right), \quad S^q_{nm}\left(\rho,\phi,z\right)=\sin\left(m\phi\right)S_{nm}\left(\rho,z\right),
\end{equation}
where,
\begin{align} \label{eq.sc1}
     C^1_{np}(k,z)=i\mu_0\rho_cnL_cke^{ik(z-pL_s)}\left(\frac{e^{-(-1)^pikL_2}+(-1)^{n+1}e^{-(-1)^pikL_1}}{n^2\pi^2-L_c^2k^2}\right),
\end{align}
\begin{align} \label{eq.sc2}
C^2_{np}(k,z)=\frac{i(-1)^pL_ck}{n\pi}C^1_{np}(k,z),
\end{align}
\begin{align} \label{eq.sc3}
C^3_{np}(k,z)=-\frac{i(-1)^pL_c|k|}{n\pi k\rho}C^1_{np}(k,z),
\end{align}
\begin{align} \label{eq.sc4}
C^4_{np}(k,z)=\frac{i|k|}{k}C^1_{np}(k,z),
\end{align}
\begin{align} \label{eq.sc5}
C^5_{np}(k,z)=-\frac{(-1)^pL_c|k|}{n\pi}C^1_{np}(k,z).
\end{align}
Performing the integrals over $k$ by splitting up the odd and even terms in $p$ while expressing the summation over the infinite pseudo-current densities through a Fourier series expansion   
\begin{align} \label{eq.diracyboy1}
    \sum_{p=-\infty}^\infty  e^{2ikpL_s} &= \frac{\pi}{L_s} \sum_{p=-\infty}^\infty \delta\left(k-\frac{\pi p}{L_s}\right),
\end{align}
we can simplify the expressions in (\ref{eq.fn}), (\ref{eq.gn}), (\ref{eq.hn}), (\ref{eq.dn}), and (\ref{eq.sn}), to find that
\begin{align}\label{eq.fnn}
    F_{n}\left(\rho,z\right)=\sum_{p=-\infty}^{\infty} I'_0\left(\left|\frac{\pi p}{L_s}\right|\rho\right)\vast(K'_{0}\left(\left|\frac{\pi p}{L_s}\right|\rho_c\right) -\frac{I'_{0}\left(\left|\frac{\pi p}{L_s}\right|\rho_c\right)K_{0}\left(\left|\frac{\pi p}{L_s}\right|\rho_s\right)}{I_{0}\left(\left|\frac{\pi p}{L_s}\right|\rho_s\right)}\vast)C^1_{np}(z),
\end{align}
\begin{align}\label{eq.gnn}
    G_{nm}\left(\rho,z\right)=\sum_{p=-\infty}^{\infty} I'_m\left(\left|\frac{\pi p}{L_s}\right|\rho\right)\vast(K'_{m}\left(\left|\frac{\pi p}{L_s}\right|\rho_c\right)-\frac{I'_{m}\left(\left|\frac{\pi p}{L_s}\right|\rho_c\right)K_{m}\left(\left|\frac{\pi p}{L_s}\right|\rho_s\right)}{I_{m}\left(\left|\frac{\pi p}{L_s}\right|\rho_s\right)}\vast)C^2_{np}(z),
\end{align}
\begin{align}\label{eq.hnn}
    H_{nm}\left(\rho,z\right)=\sum_{p=-\infty}^{\infty} mI_m\left(\left|\frac{\pi p}{L_s}\right|\rho\right)\vast(K'_{m}\left(\left|\frac{\pi p}{L_s}\right|\rho_c\right)-\frac{I'_{m}\left(\left|\frac{\pi p}{L_s}\right|\rho_c\right)K_{m}\left(\left|\frac{\pi p}{L_s}\right|\rho_s\right)}{I_{m}\left(\left|\frac{\pi p}{L_s}\right|\rho_s\right)}\vast)C^3_{np}(z),
\end{align}
\begin{align}\label{eq.dnn}
    D_{n}\left(\rho,z\right)=\sum_{p=-\infty}^{\infty} I_0\left(\left|\frac{\pi p}{L_s}\right|\rho\right)\vast(K'_{0}\left(\left|\frac{\pi p}{L_s}\right|\rho_c\right)-\frac{I'_{0}\left(\left|\frac{\pi p}{L_s}\right|\rho_c\right)K_{0}\left(\left|\frac{\pi p}{L_s}\right|\rho_s\right)}{I_{0}\left(\left|\frac{\pi p}{L_s}\right|\rho_s\right)}\vast)C^4_{np}(z),
\end{align}
\begin{align}\label{eq.snn}
    S_{nm}\left(\rho,z\right)=\sum_{p=-\infty}^{\infty} I_m\left(\left|\frac{\pi p}{L_s}\right|\rho\right)\vast(K'_{m}\left(\left|\frac{\pi p}{L_s}\right|\rho_c\right)-\frac{I'_{m}\left(\left|\frac{\pi p}{L_s}\right|\rho_c\right)K_{m}\left(\left|\frac{\pi p}{L_s}\right|\rho_s\right)}{I_{m}\left(\left|\frac{\pi p}{L_s}\right|\rho_s\right)}\vast)C^5_{np}(z),
\end{align}
where,
\begin{align} \nonumber
     C^1_{np}(z)=i\mu_0\rho_cnL_c\frac{pe^{i\frac{\pi p z}{L_s}}}{n^2L_s^2-L_c^2p^2} &\Bigg(\left((-1)^pe^{i\frac{\pi p L_2}{L_s}} +e^{-i\frac{\pi p L_2}{L_s}}\right) \\ &\qquad\qquad +(-1)^{n+1}\left((-1)^pe^{i\frac{\pi p L_1}{L_s}}+e^{-i\frac{\pi p L_1}{L_s}}\right)\Bigg),
\end{align}
\begin{align} \nonumber
C^2_{np}(z)=\frac{\mu_0\rho_cL_c^2}{L_s}\frac{p^2e^{i\frac{\pi p z}{L_s}}}{n^2L_s^2-L_c^2p^2} &\Bigg(\left((-1)^pe^{i\frac{\pi p L_2}{L_s}} -e^{-i\frac{\pi p L_2}{L_s}}\right) \\ &\qquad\qquad +(-1)^{n+1}\left((-1)^pe^{i\frac{\pi p L_1}{L_s}}-e^{-i\frac{\pi p L_1}{L_s}}\right)\Bigg),
\end{align}
\begin{align} 
C^3_{np}(z)=-\frac{|p|L_s}{p^2\pi\rho}C^2_{np}(z),
\end{align}
\begin{align} 
C^4_{np}(z)=\frac{i|p|}{p}C^1_{np}(z),
\end{align}
\begin{align} 
C^5_{np}(z)=\frac{i|p|}{p}C^2_{np}(z).
\end{align}

\newpage

\section*{Supplementary Material}
\subsection*{A solenoidal coil}
As shown in previous work~\cite{Solenoid1, solenoid2, doi:10.1063/1.1719514}, a solenoid of the same length as the high-permeability cylinder provides the most optimal solution for generating a constant axial field. Due to the image currents, the finite solenoid effectively acts as one of infinite extension, resulting in the most homogeneous possible magnetic field in the $z$-direction. A perfect finite solenoid of length $L_c$ carries a total current of 
\begin{equation}
    I_t=\int_{-L_c/2}^{L_c/2} \mathrm{d}z' \ I,
\end{equation}
where $I$ is the current density, i.e. the current per unit solenoid length, resulting in the azimuthal current $I_t/L_c$. Using (\ref{eq.fp}) the Fourier transform of the current through a finite solenoid given by
\begin{equation}
    J_\phi^{mp}(k)=I_t\delta_{m0} e^{-ikpL_c}\textnormal{Sinc}(kL_c/2).
\end{equation}
Substituting the above equation into \eqref{eq.Bzc} gives the magnetic field in the $z$-direction
\begin{align}
    B_z\left(\rho,\phi,z\right)=\frac{\mu_0I_t\rho_c}{2\pi }\sum_{p=-\infty}^{\infty}\int_{-\infty}^{\infty}\mathrm{d}k \ |k| e^{ikz}e^{-ikpL}\textnormal{Sinc}(kL_c/2)I_{0}(|k|\rho) \nonumber \qquad\qquad\qquad\qquad\\ \times\Bigg[K_{1}(|k|\rho_c)+\frac{I_{1}(|k|\rho_c)K_{0}(|k|\rho_s)}{I_{0}(|k|\rho_s)}\Bigg].
\end{align}
Performing the integral over $k$ by expressing the summation over the infinite pseudo-current reflections through a Fourier series expansion
\begin{align} \label{eq.diracyboy}
    \sum_{p=-\infty}^\infty  e^{ikpL_s} &= \frac{2\pi}{L_s} \sum_{p=-\infty}^\infty \delta\left(k-\frac{2\pi p}{L_s}\right),
\end{align}
we find
\begin{align}
    B_z\left(\rho,\phi,z\right)=\frac{\mu_0I_t\rho_c}{L_c}\sum_{p=-\infty}^{\infty}\left|\frac{2\pi p}{L_c}\right|\cos\left(\frac{2\pi p z}{L_c}\right) \textnormal{Sinc}(\pi p)I_{0}\left(\left|\frac{2\pi p }{L_c}\right|\rho\right) \nonumber \qquad\qquad\qquad\qquad\qquad\\ \times \Bigg[K_{1}\left(\left|\frac{2\pi p }{L_c}\right|\rho_c\right)+\frac{I_{1}(|\frac{2\pi p }{L_c}|\rho_c)K_{0}(|\frac{2\pi p }{L_c}|\rho_s)}{I_{0}(|\frac{2\pi p }{L_c}|\rho_s)}\Bigg].
\end{align}
This summation can be simplified as the only contributing term is $p=0$, which, when evaluated results in
\begin{equation}
     B_z\left(\rho,\phi,z\right)=\frac{\mu_0I_t}{L_c}.
\end{equation}
This result might seem counter intuitive because the magnetic field is identical to a long solenoid in free space with $N$ turns, i.e. $B_z\left(\rho,\phi,z\right)=\mu_0IN/L_c$, with \emph{no} field created by the cylindrical surface of the passive shield. This is, however, entirely physical as an infinite solenoid generates zero magnetic field outside of the solenoid itself. Consequently, there is no response due to the surface of the cylindrical wall and, therefore, no magnetic field generated by the cylindrical surface of the perfect magnetic conductor, i.e. the high-permeability cylinder.

\subsection*{Example coil designs}
\begin{figure*}[!htb]
     \centering
         {\includegraphics[scale=0.3185]{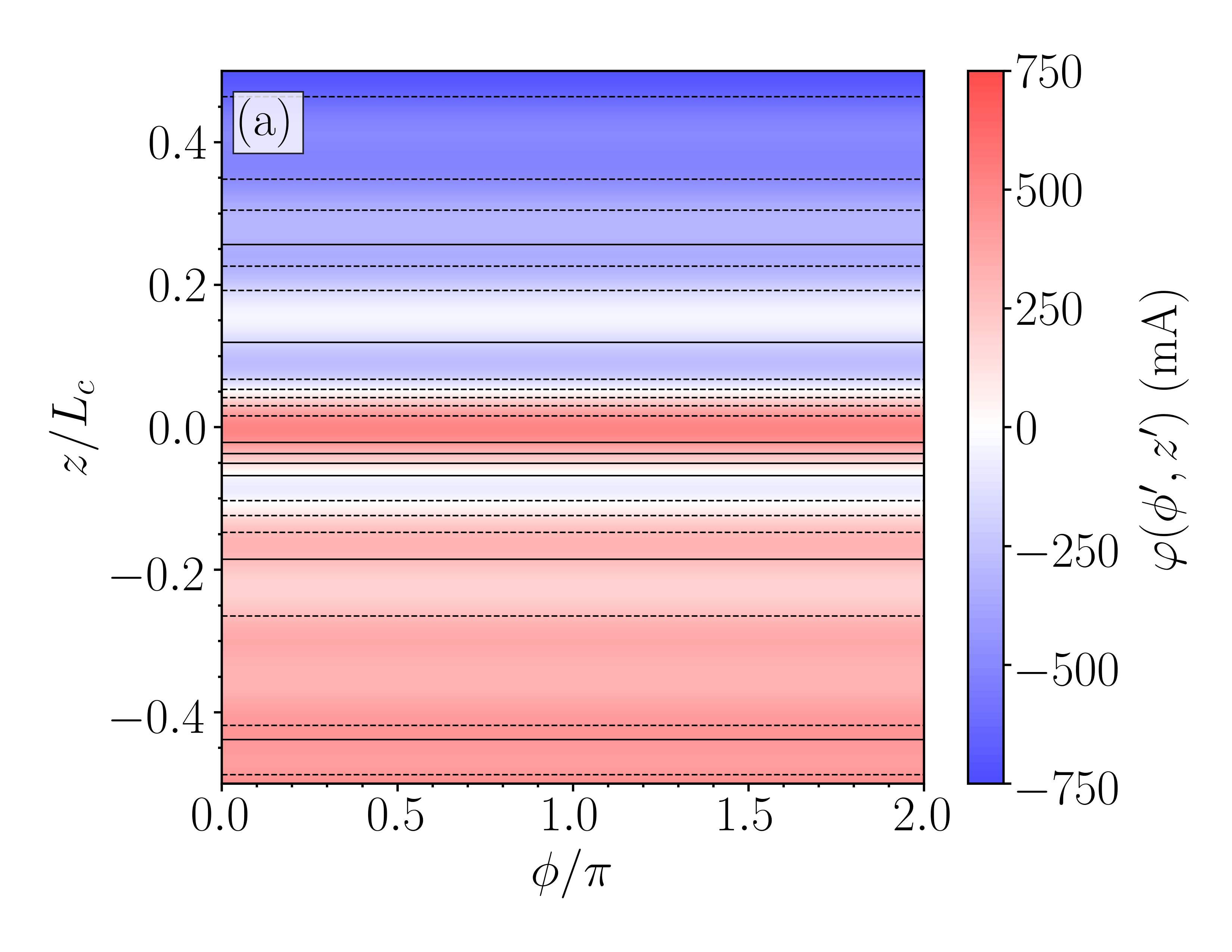}}
         {\includegraphics[scale=0.3185]{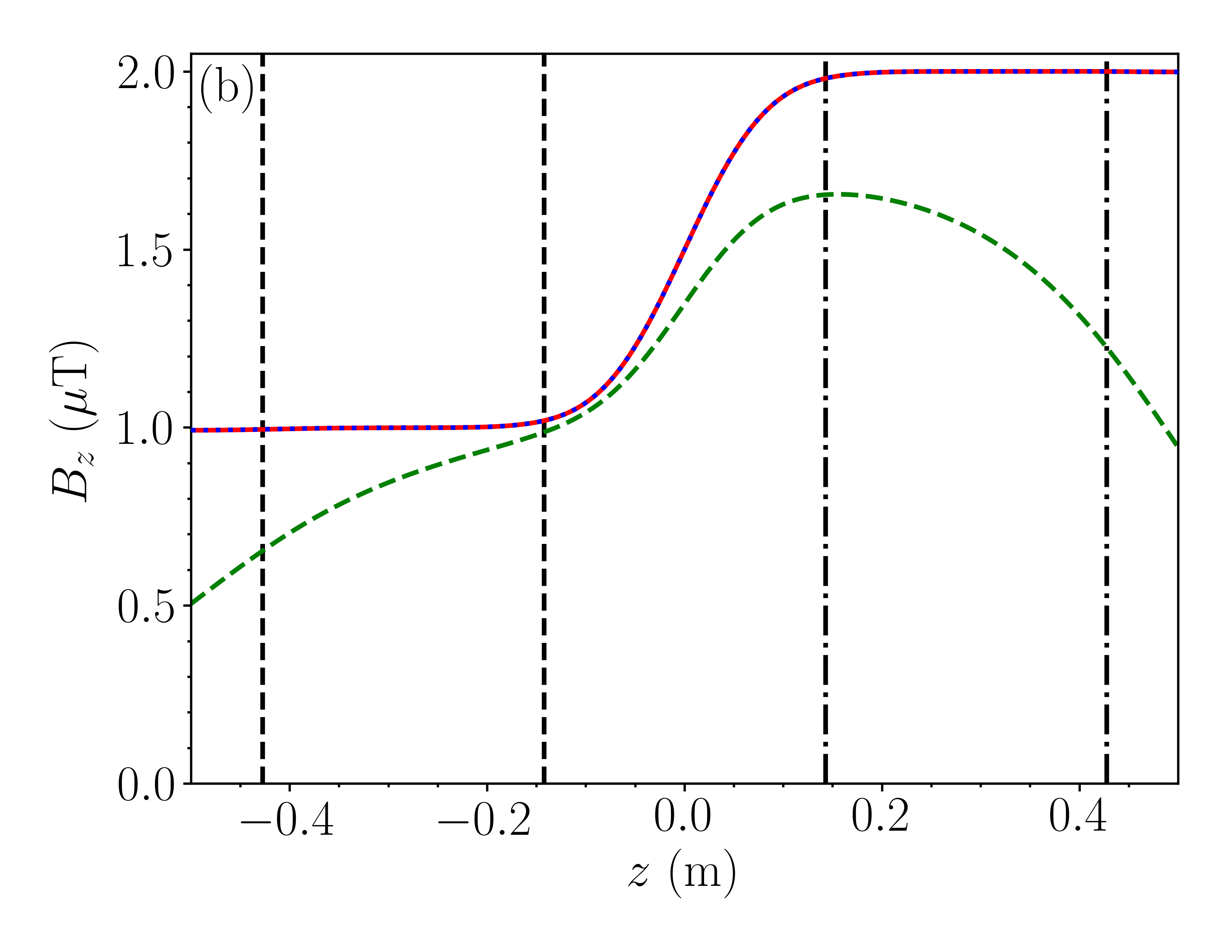}}
        \caption{Wire layouts (a) and performance (b) of an optimized hybrid active--passive uniform axial field-generating system in which current flows on a cylinder of length $L_c=0.95$ m and radius $\rho_c=0.245$ m. The wire layouts are optimized to generate axial constant fields, $B_z$, of two different magnitudes along the $z$-axis of the cylinder in two separate regions. The current-carrying cylinder is placed symmetrically inside a perfect closed magnetic shield of length $L_s=1$ m and radius $\rho_s=0.25$ m and target magnetic field values of $B_z=1$~$\mu$T and $B_z=2$~$\mu$T are optimized between $\rho=[0,\rho_c/2]$ and either $z=[-0.45L_c,0.15L_c]$ [dashed black lines in (b) where target $B_z=1$~$\mu$T] and $z=[0.15L_c,0.45L_c]$ [dot-dashed black lines in (b) where target $B_z=2$~$\mu$]. The least squares optimization was performed with parameters $N=200$, $M=0$, and $\beta=2.97\times10^{-11}$. (a) Color map of the optimal current streamfunction on the cylinder [blue and red shaded regions correspond to the flow of current in opposite senses and their intensity shows the streamfunction magnitude from low (white) to high (intense color)]. Solid and dashed black curves represent discrete wires with opposite senses of current flow, approximating the current continuum with $N_\varphi=8$ contour levels. (b) Axial magnetic field, $B_z$, versus axial position, $z$, calculated from the current continuum in (a) in three ways: analytically using \eqref{eq.brf}-\eqref{eq.bzf} (solid red curve); numerically using COMSOL Multiphysics\textsuperscript{\textregistered} Version 5.5a and modelling the high-permeability cylinder as a perfect magnetic conductor (blue dotted curve, which essentially overlays the red curve); numerically \emph{without} the high-permeability cylinder and using the Biot–Savart law with $N_{\varphi}=250$ contour levels (dashed green curve).}
        \label{fig.dualcoil}
\end{figure*}

\begin{figure*}[!htb]
     \centering
         {\includegraphics[scale=0.3185]{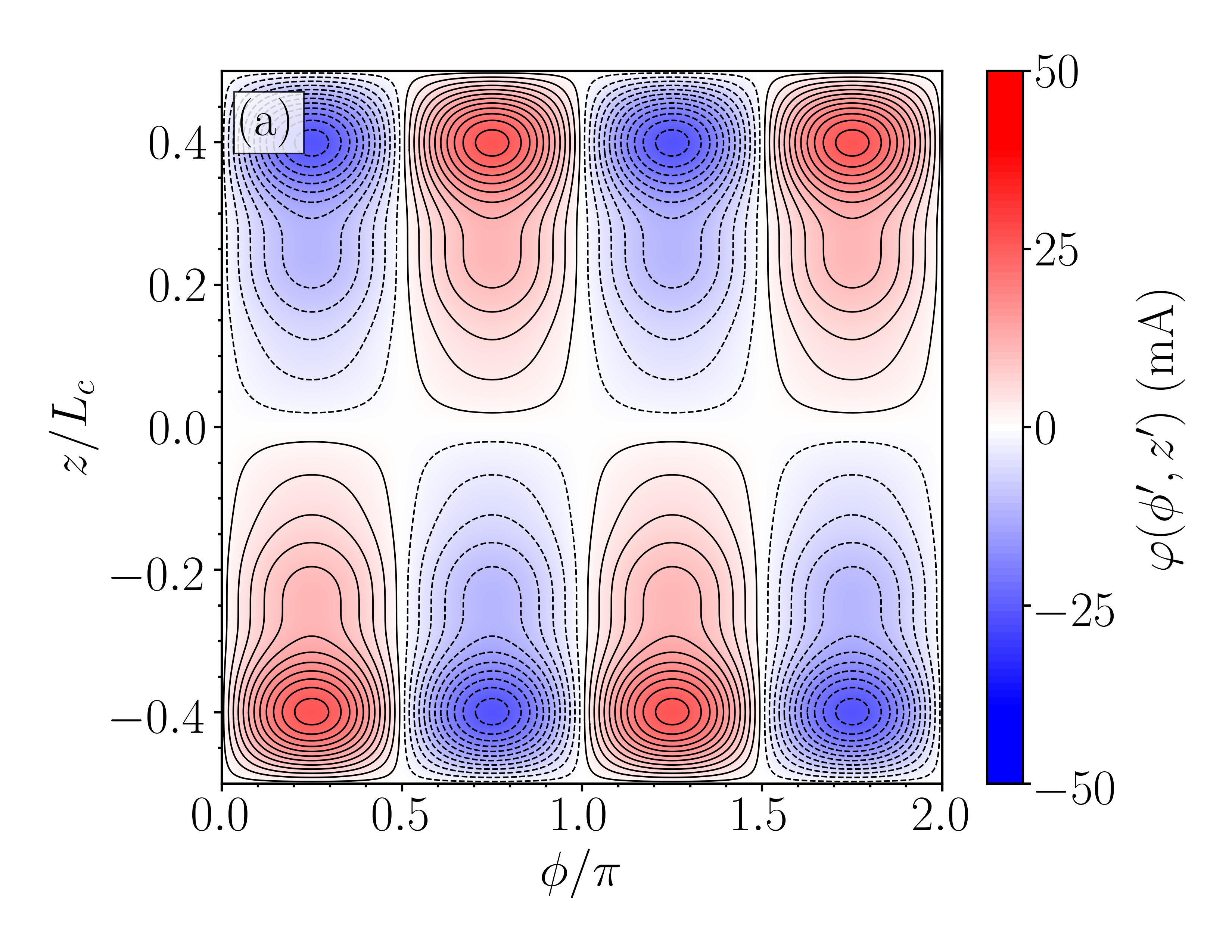}}
         {\includegraphics[scale=0.3185]{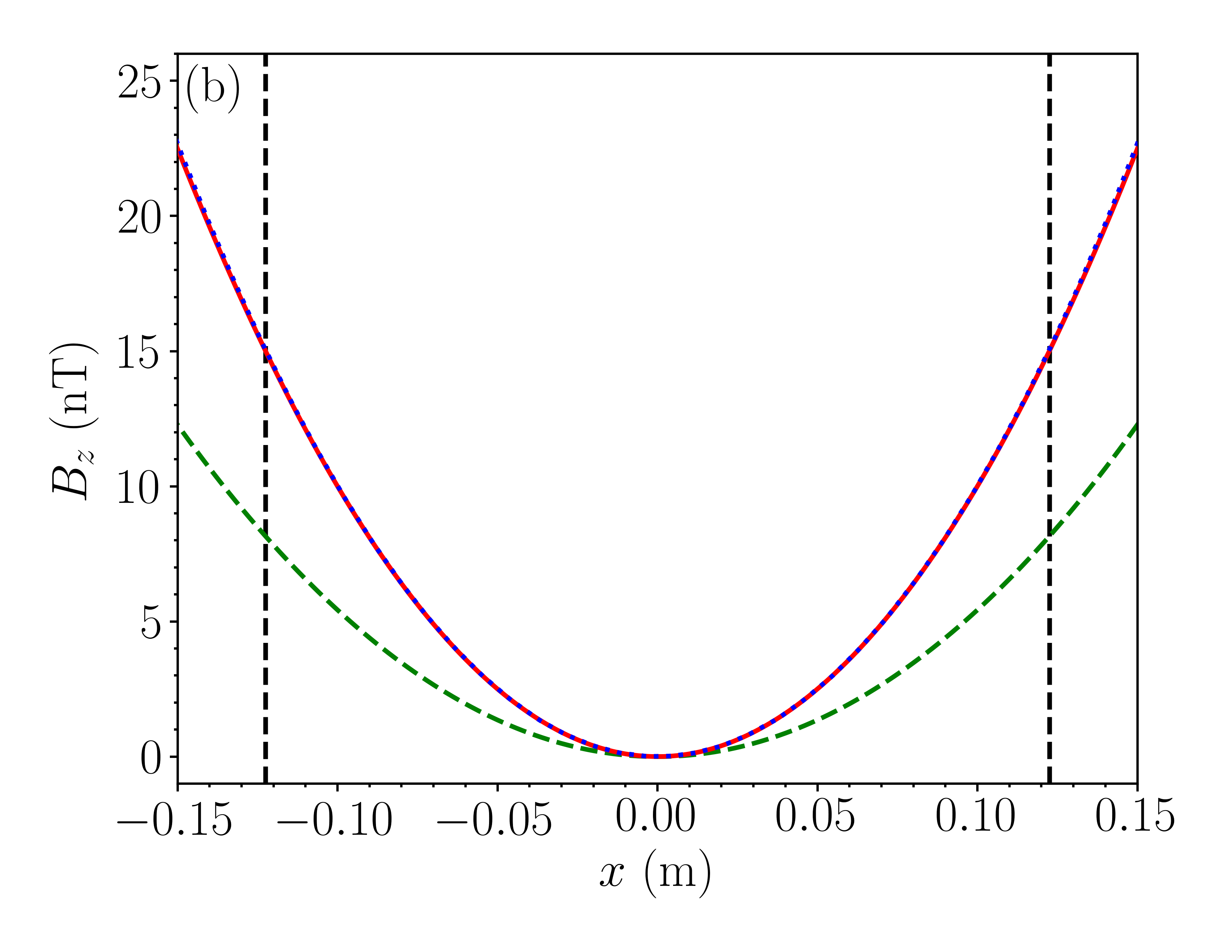}}
         {\includegraphics[scale=0.3185]{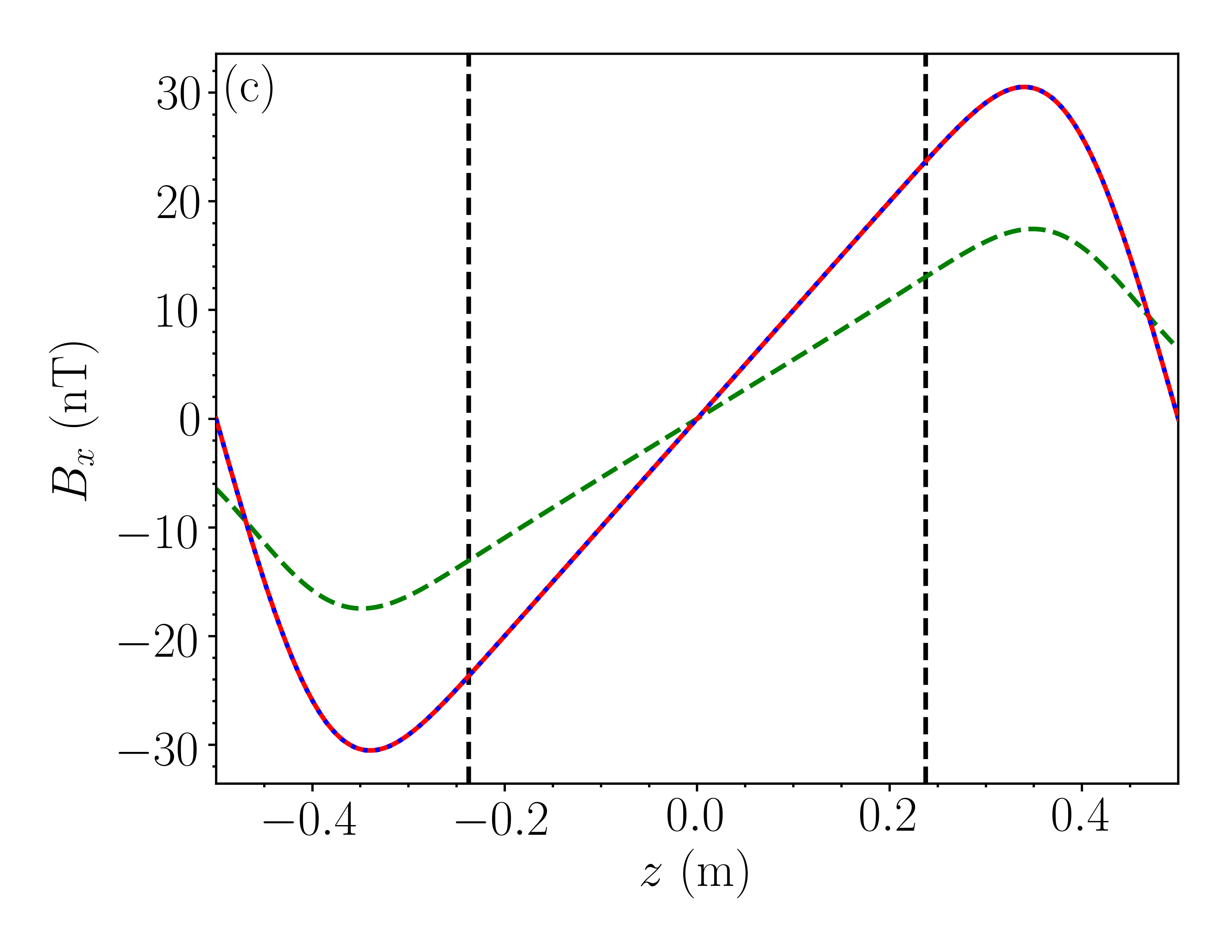}}
        \caption{Wire layouts (a) and performance (b--c) of an optimized hybrid active--passive quadratic gradient field-generating system in which current flows on a cylinder of length $L_c=0.95$ m and radius $\rho_c=0.245$ m. The wire layouts are optimized to generate a quadratic gradient field, $\mathbf{B}=(2xz~\boldsymbol{\hat{x}} - 2yz~\boldsymbol{\hat{y}} + (x^2 - y^2)~\boldsymbol{\hat{z}})$ $\mu$T. The current-carrying cylinder is placed symmetrically inside a perfect closed magnetic shield of length $L_s=1$ m and radius $\rho_s=0.25$ m and the magnetic field is optimized between $\rho=[0,\rho_c/2]$ and $z=\pm{L_c}/4$; dashed black lines in (b--c), respectively. The least squares optimization was performed with parameters $N=200$, $M=2$, and $\beta=5.95\times10^{-12}$. (a) Color map of the optimal current streamfunction on the cylinder [blue and red shaded regions correspond to the flow of current in opposite senses and their intensity shows the streamfunction magnitude from low (white) to high (intense color)]. Solid and dashed black curves represent discrete wires with opposite senses of current flow, approximating the current continuum with $N_\varphi=24$ contour levels. (b) Axial magnetic field, $B_z$, versus transverse position, $x$, at $z=0$ calculated from the current continuum in (a) in three ways: analytically using \eqref{eq.brf}-\eqref{eq.bzf} (solid red curve); numerically using COMSOL Multiphysics\textsuperscript{\textregistered} Version 5.5a and modelling the high-permeability cylinder as a perfect magnetic conductor (blue dotted curve, which essentially overlays the red curve); numerically \emph{without} the high-permeability cylinder and using the Biot–Savart law with $N_{\varphi}=250$ contour levels (dashed green curve). (c) Transverse magnetic field, $B_x$, versus axial position, $z$, at $x=0.05$ m calculated from the current continuum in (a) in the same three ways as (b); labelled as in (b).}
        \label{fig.T32coil}
\end{figure*}

In Figs.~\ref{fig.dualcoil}--\ref{fig.T32coil}, we present further hybrid active–passive systems, which generate more complex magnetic field landscapes beyond the constant field and field gradient considered in the main body of the paper. The co-ordinate axes and magnetic field plots are labelled in the same way as the systems presented in the main text. The design in Fig.~\ref{fig.dualcoil} generates uniform axial magnetic fields, $B_z$, with different strengths in two different regions of a magnetic shield. The design in Fig.~\ref{fig.T32coil} generates a quadratic gradient field whose spatial variation matches the harmonic $\mathbf{B}=(2xz~\boldsymbol{\hat{x}} - 2yz~\boldsymbol{\hat{y}} + (x^2 - y^2)~\boldsymbol{\hat{z}})$. Further systems can be designed using Python code in the GitHub repository listed in the addendum.


\end{document}